\documentclass{jpp}
\usepackage{graphicx}
\usepackage{url}
\usepackage[dvipsnames]{xcolor}

\usepackage[utf8]{inputenc}
\usepackage[T1]{fontenc}
\usepackage{amsmath,amssymb,bm,xcolor}

\shorttitle{Current sheets in relativistic weak Alfv\'{e}nic turbulence}
\shortauthor{B. Ripperda, J.F. Mahlmann, et al.}

\title{Weak Alfv\'{e}nic turbulence in relativistic plasmas II: Current sheets and dissipation}

\author{B. Ripperda\aff{1,2}
  \corresp{\email{bripperda@flatironinstitute.org}},
  J.F. Mahlmann\aff{2}\corresp{\email{mahlmann@princeton.edu}},
  A. Chernoglazov\aff{1,3},\\
  J. M. TenBarge\aff{2,4},
  E.R. Most\aff{5,6,7},
  J. Juno\aff{8},
  Y. Yuan\aff{1},\\
  A.A. Philippov\aff{1},
  \and  A. Bhattacharjee\aff{2,4}}

\affiliation{\aff{1}Center for Computational Astrophysics, Flatiron Institute, 162 Fifth Avenue, New York, NY 10010, USA
\aff{2}Department of Astrophysical Sciences, Peyton Hall, Princeton University, Princeton, NJ 08544, USA
\aff{3}Department of Physics, University of New Hampshire, 9 Library Way, Durham NH 03824, USA
\aff{4}Princeton Center for Heliophysics, Princeton University, Princeton, NJ 08540.
\aff{5}Princeton Center for Theoretical Science, Princeton University, Princeton, NJ 08544, USA
\aff{6}Princeton Gravity Initiative, Princeton University, Princeton, NJ 08544, USA
\aff{7}School of Natural Sciences, Institute for Advanced Study, Princeton, NJ 08540, USA
\aff{8}Department of Physics and Astronomy, University of Iowa, Iowa City IA 52242, USA}

\begin{document}

\maketitle

\begin{abstract}
Alfv\'{e}n waves as excited in black hole accretion disks and neutron star magnetospheres are the building blocks of turbulence in relativistic, magnetized plasmas. A large reservoir of magnetic energy is available in  these systems, such that the plasma can be heated significantly even in the weak turbulence regime. We perform high-resolution three-dimensional simulations of counter-propagating Alfv\'{e}n waves, showing that an $E_{B_{\perp}}(k_{\perp}) \propto k_{\perp}^{-2}$ energy spectrum develops as a result of the weak turbulence cascade in relativistic magnetohydrodynamics and its infinitely magnetized (force-free) limit. The plasma turbulence ubiquitously generates current sheets, which act as locations where magnetic energy dissipates. We show that current sheets form as a natural result of nonlinear interactions between counter-propagating Alfv\'{e}n waves. These current sheets form due to the compression of elongated eddies, driven by the shear induced by growing higher order modes, and undergo a thinning process until they break-up into small-scale turbulent structures. We explore the formation of {current sheets} both in overlapping waves and in localized wave packet collisions. The relativistic interaction of localized Alfv\'{e}n waves induces both Alfv\'{e}n waves and fast waves and efficiently mediates the conversion and dissipation of electromagnetic energy in astrophysical systems. Plasma energization through reconnection in current sheets emerging during the interaction of Alfv\'{e}n waves can potentially explain X-ray emission in black hole accretion coronae and neutron star magnetospheres.
\end{abstract}

\section{Introduction}

Plasma turbulence is ubiquitous in the Universe and it is not fully understood how energy cascades to small scales, and how it is eventually dissipated. Turbulent plasma in the magnetospheres of compact objects is typically magnetically dominated, characterized by a magnetization parameter $\sigma = B^2 / (4\pi \rho c^2) \geq 1$, indicating that the magnetic energy density $B^2/(4\pi)$ is larger than the rest-mass energy density $\rho c^2$ {(\citealt{Goldreich1969,Blandford1977,Duncan1992}). The large reservoir of magnetic energy in such relativistic plasmas can be transferred to kinetic and thermal energy through a turbulent cascade. In the relativistic regime, even low-amplitude waves can interact to liberate a significant amount of magnetic energy for dissipation.} {A variety of astrophysical systems rely on understanding the long-term dynamics of wave interactions, namely how they drive turbulent energy cascades and how energy is eventually dissipated at the smallest scales. Specifically, they are important to understand plasma dynamics in neutron star magnetospheres and X-ray emitting coronae around black hole accretion disks.
Magnetars, for example, can emit initially low-amplitude Alfv\'{e}n waves through star quakes that interact with the highly magnetized magnetosphere (\citealt{Li_2019,Bransgrove_2020,Yuan_2020}). Turbulent black hole accretion disks can emit similar waves, propagating into their magnetized coronae, that can interact with counter-propagating waves (\citealt{Thompson_1998,Chandran_2018}). In this work, we explore how overlapping, low-amplitude, Alfv\'{e}n waves interact in the highly magnetized plasma regime, how they create a weak turbulence spectrum, and eventually dissipate.}
{In more realistic magnetospheric settings, Alfv\'{e}nic fluctuations can be localized in wave-packets that travel along a strong guide field \citep[see, e.g.,][]{Li_2019,yuan2020alfvn}. Besides magnetic energy dissipation due to the interaction of wave packets, it is crucial to understand the production and dynamics of fast waves. Unlike Alfv\'{e}n waves, fast modes are not confined to travel along the strong guide field and are, thus, suitable candidates to transport electromagnetic energy out of the magnetosphere.}

{We explore the development of weak turbulence through wave interactions in the relativistic, highly magnetized, magnetohydrodynamics (MHD) regime.} {While MHD is a valid theory for high magnetization, numerically it is difficult to accurately describe highly magnetized plasma at $\sigma \gtrsim \mathcal{O}(10^2)$ (\citealt{Noble2006,Ripperda_2019}). Compact object magnetospheres in particular can be so highly magnetized that the infinitely magnetized, or force-free, limit of relativistic MHD\footnote{{Note that the Newtonian force-free limit is a subset of the relativistic force-free MHD equations. We refer the reader to the introduction of \citet{MahlmannPhD_2020} for a detailed review of and context for relativistic force-free methods in astrophysics.}} is not only applicable (\citealt{Goldreich1969,Blandford1977}) but also provides a more accurate numerical method. Although the force-free MHD limit technically cannot describe the physical effects of dissipation, it does capture the formation of a turbulent cascade (\citealt{Thompson_1998,Cho_2005,Li_2019}) and of current sheets (\citealt{mahlmann2020computational}).}
The force-free MHD limit admits two normal modes, an Alfv\'{e}n wave and a fast wave. Nonlinear interactions of these waves are expected to result in a turbulent cascade {\citep[cf.][in the ideal relativistic MHD limit]{Takamoto_2017}}. Wave interactions in the force-free MHD limit have been studied analytically and numerically (\citealt{Thompson_1998,Heyl_1999,Troischt_2004,Cho_2005,Li_2019}), emphasizing the dominance of either a four-wave interaction of two incoming and two outgoing Alfv\'{e}n waves as the building blocks of relativistic MHD turbulence \citep[{following the Newtonian theory of}][]{sridhar1994}, or three-wave interaction of two colliding Alfv\'{e}n waves producing a fast wave; two fast waves producing a fast wave; and an Alfv\'{e}n and a fast wave producing an Alfv\'{e}n wave. Newtonian theory has, however, firmly shown that Alfv\'{e}n-Alfv\'{e}n three-wave interactions, resulting in an outgoing Alfv\'{e}n wave, are the dominant building block for MHD turbulence (\citealt{Montgomery1995,Ng1996,Howes_2013,Nielson_2013}). {\citet[][Paper I of this sequence]{Tenbarge2021} concluded} both analytically and numerically that the three-wave interaction of two Alfv\'{e}n waves resulting in a tertiary Alfv\'{e}n mode is also the fundamental physical process underlying weak \emph{relativistic} MHD turbulence. 

{The slope of the turbulent energy spectrum that results from the wave interactions is} an open issue that has crucial implications for both plasma turbulence theory {as well as} {plasma dynamics in compact object magnetospheres.} In the presence of a uniform magnetic field, the energy spectrum that develops from two interacting Alfv\'{e}n waves is expected to be anisotropic based on Newtonian MHD weak turbulence theory. In other words, the energy cascade occurs entirely in the direction perpendicular ($\perp$) to the uniform field. 
In this case, one expects an energy spectrum $E_{B_\perp}(k_{\perp}) \propto k_{\perp}^{-2}$ to develop, where $k_{\perp}$ is the perpendicular Fourier component of the waves (\citealt{Ng1997,Galtier00,Bhattacharjee_2001,Kuznetsov_2001}). In Paper I we {argue} that this prediction holds in relativistic MHD turbulence, very much akin to Newtonian theory. {In Paper II, we study the shape and formation of a turbulent spectrum due to the long-term interaction of initially low-amplitude Alfv\'{e}n waves.}

{In the Newtonian limit, it has been shown that} plasma turbulence intermittently generates small-scale coherent structures in the form of current sheets, providing the main locations where magnetic energy is dissipated after it cascades (\citealt{Matthaeus:1986,TenBarge_2013,Zhdankin_2013,Howes_2016,dong2018,Verniero_2018,Verniero_2018b,rueda2021threedimensional}). These current sheets can be viewed as small-scale eddies stretched along the magnetic field lines, which form as the cascade proceeds towards the smallest dissipative scales (\citealt{Boldyrev_2006,Mallet:2017,Loureiro:2017,comisso2018}). {Recently, such coherent structures were confirmed to also form in relativistic plasma turbulence (\citealt{Zhdankin_2017,Comisso_2018,nattila2020radiative}) and {general relativistic,} turbulent, black hole accretion flows (\citealt{Nathanail_2020,Ripperda_2020}). Here, we explore the formation of current sheets as a result of fundamental wave interactions in relativistic, highly magnetized plasma, and their effect on the turbulent spectrum.}

This work is organized as a sequence of papers. This manuscript (Paper II) extends the interaction of counter-propagating Alfv\'{e}n waves (as examined asymptotically in Paper I) up to their turbulent decay in the far nonlinear regime by numerically solving the full three-dimensional (3D) set of special relativistic ideal MHD equations and their force-free limit. We compare results from independent and different numerical methods to substantiate our findings. The algorithms we employ{, the high-order force-free code \texttt{ET-FFE} (\citealt{mahlmann2020computationala}) and the force-free/relativistic MHD code $\texttt{BHAC}$ (\citealt{BHAC,Olivares2019,Ripperda_2019}),} are described in Section \ref{sec:numerics}.
We explore the nonlinear modeling of continuously overlapping Alfv\'{e}n waves in detail in Section \ref{sec:alfvendynamics}, {where we provide an analysis of current sheet formation in developing weak turbulence}. The formation of current sheets and the dynamics of fast modes as a result of the interaction between localized Alfv\'{e}n wave packets are examined in Section \ref{sec:alfvenpackets}. {We conclude with a discussion of the obtained results in Section \ref{sec:conclusions}.}

\section{Numerical methods}
\label{sec:numerics}
We solve the set of 3D ideal special-relativistic MHD equations in Cartesian coordinates $(x,y,z)$ {using Lorentz-Heaviside units, where a factor of $1/\sqrt{4\pi}$ is absorbed into the electromagnetic fields and velocities are measured in units of the speed of light $c=1$}:
\begin{equation}
\partial_t\bm{B} + \nabla \times \bm{E} = 0,
\label{eq:faraday4}
\end{equation}
\begin{equation}
\partial_t D + \nabla \cdot \left(\rho \Gamma \bm{v}\right) = 0,
\label{eq:numberdensity}
\end{equation}
\begin{equation}
\partial_t\tau + \nabla \cdot \left(\bm{E}\times\bm{B} + \left(h \Gamma^2 -D\right) \bm{v}\right) = 0,
\label{eq:energy}
\end{equation}
\begin{equation}
\partial_t\bm{S} + \nabla \cdot \left(-\bm{E}\bm{E} - \bm{B}\bm{B} + h \Gamma^2 \bm{v}\bm{v} + \left[\frac{1}{2}\left(E^2+B^2\right) + p\right] \right) = 0.
\label{eq:momentum}
\end{equation}
{In the ideal MHD limit, the electric field is not evolved, but given by the relation $\bm{E} = -\bm{v}\times\bm{B}$, dependent on the velocity field $\bm{v}$ and the magnetic field $\bm{B}$. The conserved mass density, $D$, energy density $\tau$, and energy flux density, $\bm{S}$, are given by}
\begin{equation}
D = \Gamma \rho,
\label{eq:densityD}
\end{equation}
\begin{equation}
\tau = \frac{1}{2}(E^2 +B^2) + h\Gamma^2 - p - D,
\label{eq:energytau}
\end{equation}
\begin{equation}
\mathbf{S} = \bm{E}\times\bm{B} + h \Gamma^2 \bm{v},
\label{eq:momentumS}
\end{equation}
where the enthalpy $h = \rho \left(1+\varepsilon\right) + p $ is given in terms of the rest-mass density $\rho$ and specific internal energy $\varepsilon$. In the following, we adopt the ideal fluid equation of state, $p= \rho \varepsilon \left(\hat{\gamma} -1\right)$, with an adiabatic index $\hat{\gamma} = 4/3$. We find it useful to define the dimensionless magnetization parameter as {the ratio between magnetic enthalpy density and rest-mass density with} $\sigma = B^2/\rho$. {We note that} $\sigma_{\rm hot} = B^2/h$ {is commonly used to specify the magnetization of hot plasma as it accounts for non-vanishing thermal pressure; it defines the Alfv\'{e}n speed $v_A= \sqrt{\sigma_{\rm hot} / \left( 1 + \sigma_{\rm hot}\right)}$}. {The gas-to-magnetic-pressure ratio $\beta=2p/B^2$ is limited to $\beta \leq 1/(2\sigma_{\rm hot})$.} The fluid velocity $\bm{v}$ measured by an inertial observer has the corresponding Lorentz factor $\Gamma = (1-v^2)^{-1/2}$. We note that an observer locally comoving with velocity $\bm{v}$ will measure the magnetic energy density $b^2 = B^2/\Gamma^2 + \left(\bm{B}\cdot\bm{v}\right)^2$.

In addition, we will also consider the limit of negligible plasma inertia and pressure, namely $\sigma\rightarrow\infty$ and, hence, $v_A \rightarrow 1$.
This limit corresponds to a vanishing Lorentz force, $q \bm{E} + \bm{J} \times \bm{B} = 0$, which implies that $\bm{E}\cdot \bm{B} =0$ and also requires magnetic dominance $E^2 < B^2$.
Here, $q = \nabla \cdot \bm{E}$ is the charge density and $\bm{J}$ is the current as seen by an inertial observer.
We treat this limit by solving the set of special-relativistic force-free MHD equations (e.g., \citealt{gruzinov1999stability,2002luml.conf..381B}):
\begin{equation}
\partial_t\bm{B} + \nabla \times \bm{E} = 0,
\label{eq:faradayFF}
\end{equation}
\begin{equation}
\partial_t\bm{E} - \nabla \times \bm{B} = -\bm{J},
\label{eq:ampereFF}
\end{equation}
with force-free current density
\begin{align}
    \bm{J} = q \frac{\bm{E}\times\bm{B}}{B^2} + \frac{\bm{B}\cdot\left(\nabla\times\bm{B}\right)-\bm{E}\cdot\left(\nabla\times\bm{E}\right)}{B^2}\bm{B}
    \label{eq:ffcurrent}
\end{align}

The force-free limit of MHD does not describe the physics of dissipative mechanisms like magnetic reconnection. Dissipation is instead the result of the numerical procedure applied to obey the ideal force-free conditions $\mathbf{E}\cdot\mathbf{B}=0$ and $E^2<B^2$. Enforcing these conditions, particularly in reconnecting current sheets, i.e., where the reconnecting components of the magnetic field go to zero, results in the removal of electromagnetic energy from the system. We perform our force-free MHD simulations with two different and independent algorithms {({\tt BHAC} and {\tt ET-FFE}, described below)} that enforce the ideal force-free conditions in different manners. We compare the results to assure that the turbulent cascade, the formation process of current sheets in weak turbulence conditions, and the thinning of those current sheets is accurately captured.
We also test the robustness and convergence of our force-free MHD results for large number of characteristic time scales by simulating the same setup for increasing resolutions. {Finally, the relativistic ideal MHD capacities of {\tt BHAC} allow us to extend our results to finite magnetizations as opposed to the force-free limit.} We outline the details of the numerical algorithms below.

\citet{mahlmann2020computationala} introduced a scheme based on the infrastructure of the \textsc{Einstein Toolkit}\footnote{\url{http://www.einsteintoolkit.org}}, here dubbed {\tt ET-FFE}, specifically designed for the high-order conservative modeling of the force-free electrodynamics equations (\ref{eq:faradayFF}-\ref{eq:ampereFF}). The charge density is evolved in a separate continuity equation and the algorithm relies on the fully consistent force-free current (\ref{eq:ffcurrent}). Hyperbolic/parabolic cleaning is utilized to maintain a solenoidal magnetic field and a conserved electric charge. 
Algebraic corrections are applied when numerical violations of the force-free constraints occur. The combination of algebraic corrections and the consistent ideal force-free current does not require implicit steps in the time-integrator. Therefore, employing MP7 \citep{Suresh1997} spatial reconstruction and fourth-order accurate time integration results in very competitive convergence of the numerical diffusion and dispersion \citep{mahlmann2020computational}. These properties are essential for the robustness of the results presented in this manuscript.

{{\tt BHAC} (\citealt{BHAC,Olivares2019}) can} capture the transformation of electromagnetic to kinetic and thermal energy {when performing} simulations for finite magnetization by evolving the full set of ideal special-relativistic MHD equations (\ref{eq:faraday4}-\ref{eq:momentum}). Additionally, the set of force-free MHD equations (\ref{eq:faradayFF}-\ref{eq:ampereFF}) is implemented in the relativistic resistive MHD framework in {\tt BHAC} \citep{Ripperda_2019,Ripperda_2019a}, in combination with the resistive force-free Ohm's law of \cite{Alic} replacing Eq.~(\ref{eq:ffcurrent}),
\begin{equation}
\bm{J} = q \frac{\bm{E}\times\bm{B}}{B^2} + \frac{1}{\eta} \left[\left(\bm{E}\cdot\bm{B}\right)\frac{\bm{B}}{B^2} + \Theta(E^2-B^2)\frac{\bm{E}}{B^2}\right],
\label{eq:ohmslaw}
\end{equation}
which imposes the force-free conditions using a damping current on time scales set by an effective resistivity $\eta$, which is smaller than the time step $\Delta t$ of our simulation. {We employ a fiducial, small, uniform, and constant resistivity $\eta=10^{-6}$, and show that larger values have a significant effect on the large-scale damping of the total electromagnetic energy due to Ohmic heating.} The second damping term {(proportional to $E^2-B^2$ in Eq.~\ref{eq:ohmslaw})} only {needs to be} activated locally when $E^2 >B^2$, which rarely occurs in weak turbulence with a strong guide field, and we therefore make use of the Heaviside function $\Theta$. {Throughout this manuscript, {\tt BHAC} is used for both relativistic ideal MHD simulations with different magnetizations and force-free MHD simulations.}
To treat small damping time scales and, hence, stiff source terms in the electric current, we use an implicit-explicit (IMEX) Runge-Kutta time integrator
\citep{Pareschi}. We compute curl and divergence terms in the evolution equations using a second-order accurate finite-volume scheme composed of a Rusanov Riemann solver (\citealt{Rusanov}) paired with a third-order accurate monotonicity preserving reconstruction scheme (\citealt{Cada}).
The solenoidal magnetic field constraint, $\nabla \cdot \bm{B} = 0$, is enforced to machine precision by means of a staggered constrained transport scheme \citep{Evans,Olivares2019}, where the electric fluxes are computed using the upwind-constrained transport scheme \citep{Londrillo}. 
Instead of evolving the charge density $q$, it is obtained by numerically taking the divergence of the evolved electric field as $q = \nabla \cdot \bm{E}$ at every time step of the simulation. While this may lead to a small non-conservation of global charge, we have found the effect to be negligible \citep{Ripperda_2019}. The implementation of the IMEX scheme used for the solution of the force-free MHD equations is presented in \cite{Ripperda_2019,Ripperda_2019a} in the context of resistive relativistic MHD.

Our results are numerically converged between the two force-free MHD algorithms at the high resolutions we employ, and in Appendix A we show that dispersion and diffusion errors never dominate on the evolved time-scales. {The high-order reconstruction capacities of {\tt ET-FFE} compensate for a significant factor of resolution needed in lower-order methods, such as {\tt BHAC}. We present results that converge between two different implementations of the force-free limit of MHD. Different techniques to process violations of the force-free conditions can alter the global field dynamics. Comparing two methods that use the two most commonly employed correction methods (namely, algebraic resets and charge conservation in {\tt ET-FFE}, versus driving currents in {\tt BHAC} with a finite resistivity $\eta$) supports the reliability and reproducibility we claim for our results.} In Appendix B, we show that the formation of current sheets in weak turbulence with a strong guide field is captured by both force-free MHD algorithms and that ideal force-free violations are negligible until the current sheets break-up.

\section{{Overlapping Alfv\'{e}n waves}}
\label{sec:alfvendynamics}


{Interacting Alfv\'{e}n waves, for example launched into the magnetosphere of a compact object, can result in the development of a weak turbulence cascade (\citealt{Chandran_2018,Yuan_2020}). In this Section, we set up a toy problem of two overlapping perpendicularly polarized counter-propagating Alfv\'{e}n waves. The waves are initialized on top of each other and cover the entire 3D domain with periodic boundaries. This initial condition reproduces all the characteristics of the nonlinear interactions described theoretically in Paper I. We explore the development of a weak turbulence spectrum, and the formation of current sheets due to the interaction of the waves, that can provide a viable dissipation route.}

\subsection{Wave initialization}
\label{sec:waveinitialization}

We consider the nonlinear interaction between two overlapping, perpendicularly polarized Alfv\'{e}n waves that counter-propagate in a periodic 3D domain along a uniform guide field $\mathbf{B_0} = B_0 \mathbf{\hat{z}}$. The waves are described by positive constants representing the components of the wave vector that are perpendicular ($k_{\perp}$) and parallel ($k_{\parallel}$) to the guide field $\mathbf{B_0}$. We employ a cubic box with $L_{\perp}=L_x=L_y=2\pi$, $L_{\parallel} = L_z = 2\pi$, and a uniform resolution of $(N_x, N_y, N_z) = (N_{\perp}, N_{\perp}, N_{\parallel})$ cells per initial wavelength. For a scale-free definition of characteristic (wave)lengths, we prescribe $k_{\parallel}=2\pi/L_{\parallel}$, and $k_{\perp}=2\pi/L_{\perp}$. Then, we examine the waves described by the wave vectors $\mathbf{k}^{+}_1=k_{\perp} {\mathbf{\hat{x}}} - k_{\parallel} {\mathbf{\hat{z}}}$, from now on represented as $(1,0,-1)$, and $\mathbf{k}^{-}_2=k_{\perp} \hat{\mathbf{y}} + k_{\parallel} \hat{\mathbf{z}}$, or $(0,1,1)$. 
The magnetic field is initialized through a vector potential $\mathbf{A} = (-B_0 y, 0, \delta B_{\perp}[\sin({k_\perp}x+{k_\parallel}z) + \sin({k_\perp}y-{k_\parallel}z)])$, representing the initial counter-propagating Alfv\'{e}n waves with frequency $\omega_0 = k_{\parallel} v_A $ and $\hat{B} = {\delta B_{\perp}}/{B_0}$.
The electric field is initialized as $\mathbf{E} = (v_A B_y, v_A B_x,0)$ such that the velocity is equal to the drift velocity $\mathbf{v} = \mathbf{E} \times \mathbf{B}/B^2$.
We note that the overlapping Alfv\'{e}n waves, in contrast to a single Alfv\'{e}n wave, are not an exact force-free MHD equilibrium due to a small second-order violation $\mathbf{E}_{\mp} \cdot \mathbf{B}_{\pm} \neq 0$ between the fields of the two waves. These perturbations are instantaneously algebraically cut in {\tt ET-FFE} and damped on a short time-scale in {\tt BHAC}. Initially, we set a gas-to-magnetic-pressure ratio of $\beta=2p/B_0^2=0.02$ and set the magnetization $\sigma \in [10^{-2};10^{-1};1;10;100]$ (corresponding to $\sigma_{\rm hot} \in [10^{-2};10^{-1};1;7;20]$), where we vary density $\rho$ and keep a constant guide field with $B_0=1$. We fix the pressure to $p=0.01$ and employ an adiabatic index $\hat{\gamma}=4/3$ for an ideal relativistic gas. In the force-free MHD case, $\beta \rightarrow 0$, $\sigma \rightarrow \infty$, and $v_A \rightarrow 1$.

{Through the three-wave interaction, turbulence can transfer magnetic energy anisotropically from large to small scales.} The two initial waves interact to form an inherently nonlinear, purely magnetic mode, physically representing a shear in the magnetic field with wave vector $\mathbf{k}^{0}_2 = k_{\perp}(\hat{\mathbf{x}}+ \hat{\mathbf{y}})$, or $(1,1,0)$, which does not grow secularly in time but oscillates at twice the frequency of the primary waves (see Paper I). 
The interaction between this secondary mode and the primary Alfv\'{e}n modes nonlinearly generates two tertiary Alfv\'{e}n waves whose energy grows secularly in time. The $\mathbf{{k_1^{+}}}$ wave transfers energy to an Alfv\'{e}n mode with $\mathbf{k}_3^{+} = 2k_{\perp} \mathbf{\hat{x}} + k_{\perp} \mathbf{\hat{y}} - k_{\parallel} \mathbf{\hat{z}}$, or $(2,1,-1)$, and the $\mathbf{{k_1^{-}}}$ wave to $\mathbf{k}_3^{-} = k_{\perp} \mathbf{\hat{x}} + 2k_{\perp} \mathbf{\hat{y}} + k_{\parallel} \mathbf{\hat{z}}$, or $(1,2,1)$.

{To capture all physical behavior including nonlinear interactions at play in magnetized turbulence, we solve the full 3D set of equations.} The strength of the nonlinearity is characterized by $\chi = k_{\perp} v_{\perp} / (k_{\parallel} v_A)$, where $k_{\perp} = k_{\parallel}$ in the setup presented in this section, and $v_{\perp} / v_A \sim \delta B_{\perp} / B_0=10^{-1}$. This gives $\chi = 10^{-1}$, resulting in a nonlinear time of $\chi^{-2} = 100$ wave-crossings $2\pi / \omega_0$. The limit of strong turbulence occurs when the nonlinear energy transfer time scale is comparable to the wave period, as to say, when $\chi \gtrsim 1$ (\citealt{goldreich1995}).
The nonlinearities governing these dominant three-wave interactions require variations in both directions perpendicular to the equilibrium magnetic field (\citealt{Howes_2013,Howes_2014}, Paper I). In the 2D limit perpendicular to the uniform field the dominant nonlinearities governing the three-wave interactions are retained, yet the linear physics of the anisotropic cascade is absent: the linear interactions representing the propagation of the Alfv\'{e}n waves along the magnetic field are only non-zero when the parallel wave-number $k_{\parallel}$ is non-zero, which requires a (third) field-parallel dimension.

\subsection{Current sheet formation}
\label{sec:currentsheetformation}

\begin{figure}
  \centering
  \includegraphics[width=1\textwidth]{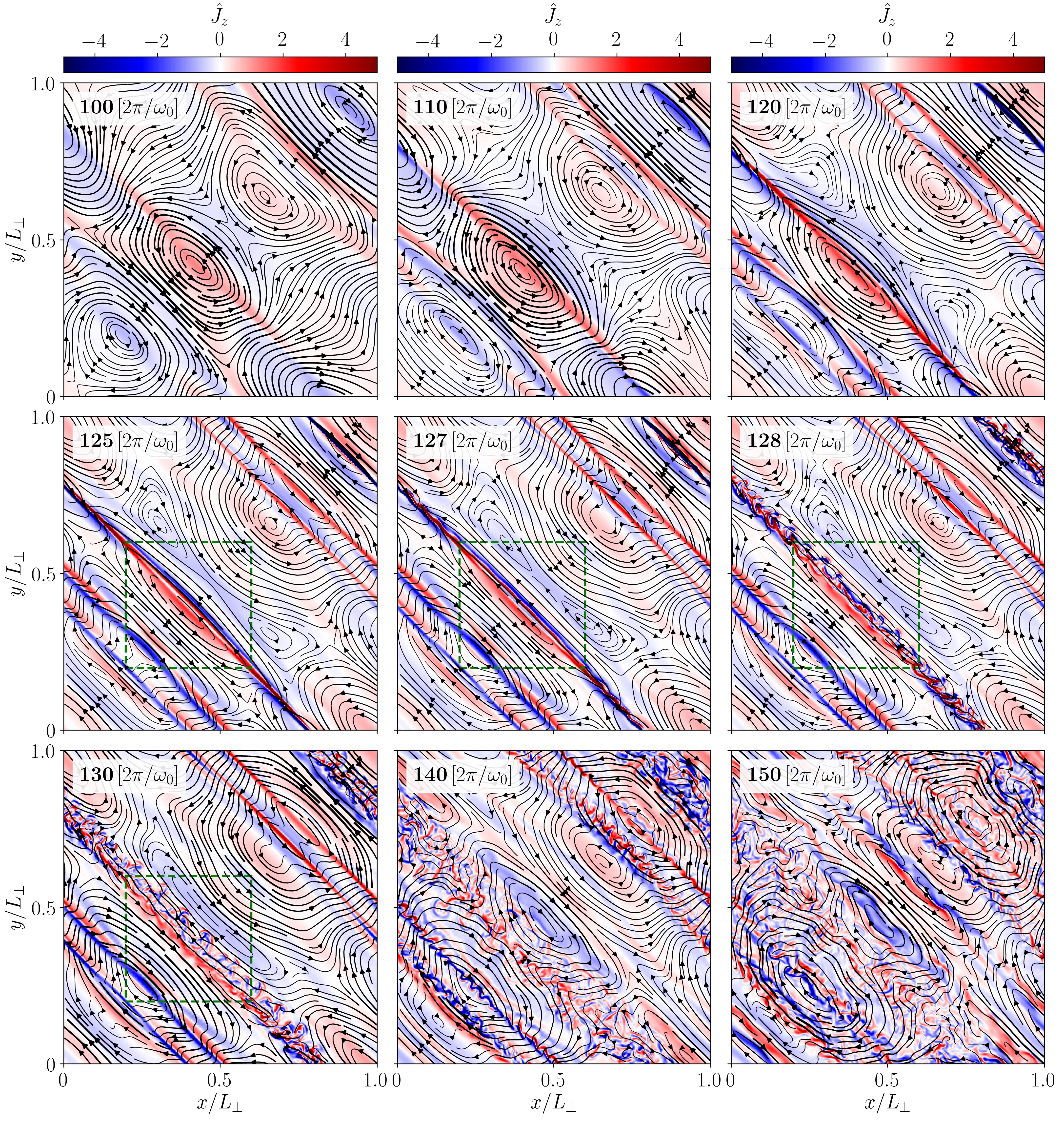}
  \vspace*{-22pt} 
    \begin{flushright}
    \textcolor{blue}{\texttt{ET-FFE}}
    \end{flushright}
  \caption{Normalized out-of-plane current density $\hat{J}_z = [\nabla \times \mathbf{B}]_z / (k_{\perp}B_0)$ in the $(x,y)$-plane perpendicular to the guide field showing the formation ($t \lesssim 100\;[2\pi/\omega_0]$), thinning ($t \sim 100-120 \;[2\pi/\omega_0]$), and break-up of current sheets ($t \sim 120-130 \;[2\pi/\omega_0]$) into developed turbulence ($t \gtrsim 130-150 \;[2\pi/\omega_0]$) with \texttt{ET-FFE}. In-plane magnetic field lines are overplotted. We note that the evolution of the current density is similar for a run at $1024^3$ cells with {\tt BHAC}. Fig.~\ref{fig:currentsheetFF512volume} presents a 3D rendering of the evolving current sheet enclosed by the dashed green rectangles. {An animated version of this figure is provided in \citet{SupplementaryMediaA}.}}
\label{fig:currentsheetFF1024}
\end{figure}

\begin{figure}
  \centering
  \includegraphics[width=0.9\textwidth]{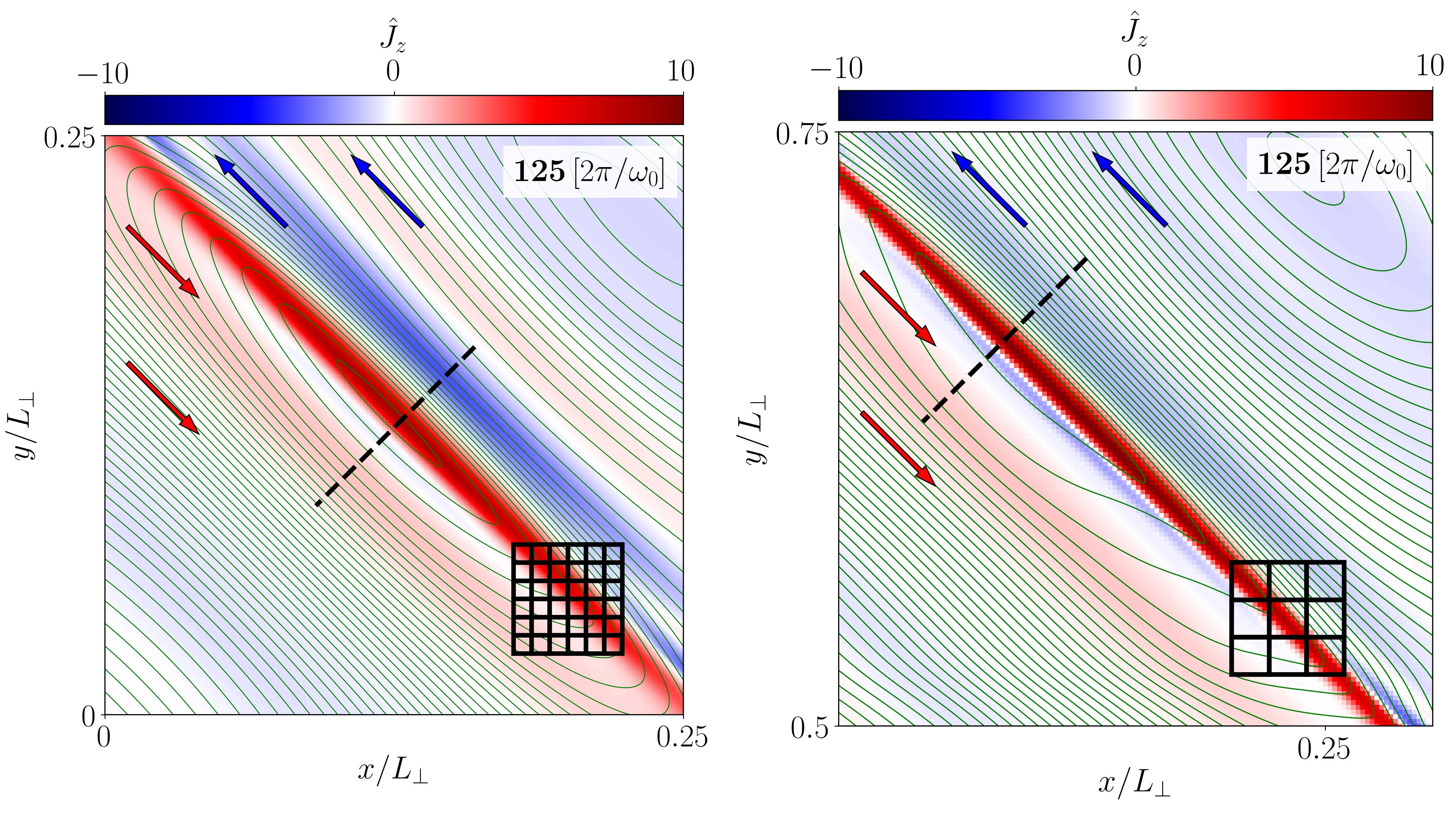}
  \vspace{-11pt}
    \begin{flushright}
    \textcolor{blue}{\texttt{BHAC}\hspace{5cm}\texttt{ET-FFE}\hspace{1cm}}
    \end{flushright}
    \caption{Current sheets form between elongated eddies. Zoomed-in view of the {out-of-plane current density $\hat{J}_z = [\nabla \times \mathbf{B}]_z / (k_{\perp}B_0)$} at 125 wave-crossings. Contours of the out-of-plane vector potential $A_z$ (where $\mathbf{B} = \nabla \times \mathbf{A}$), representing in-plane magnetic field lines, are overplotted in green, showing that regions of large current density originate from compressed magnetic field (direction indicated by arrows). We select the strongest current sheet in the domain of a \texttt{BHAC} $1024^3$ simulation (\textit{left}), and an \texttt{ET-FFE} $512^3$ setup (\textit{right}). To indicate the respective applied resolutions, we show a fraction of the numerical grid stacked with blocks, where each block consists of $8\times8$ cells. The current sheets (in bright red) are covered along their thickness $\delta \approx 0.01 \times 2\pi \approx 0.06$ by at least 8 cells in both cases. A 1D outline of field quantities across the center of each current sheet (dashed black lines) is presented in Fig.~\ref{fig:CSCompare}.}
\label{fig:Compression}
\end{figure}

\begin{figure}
  \centering
  \includegraphics[width=1\textwidth]{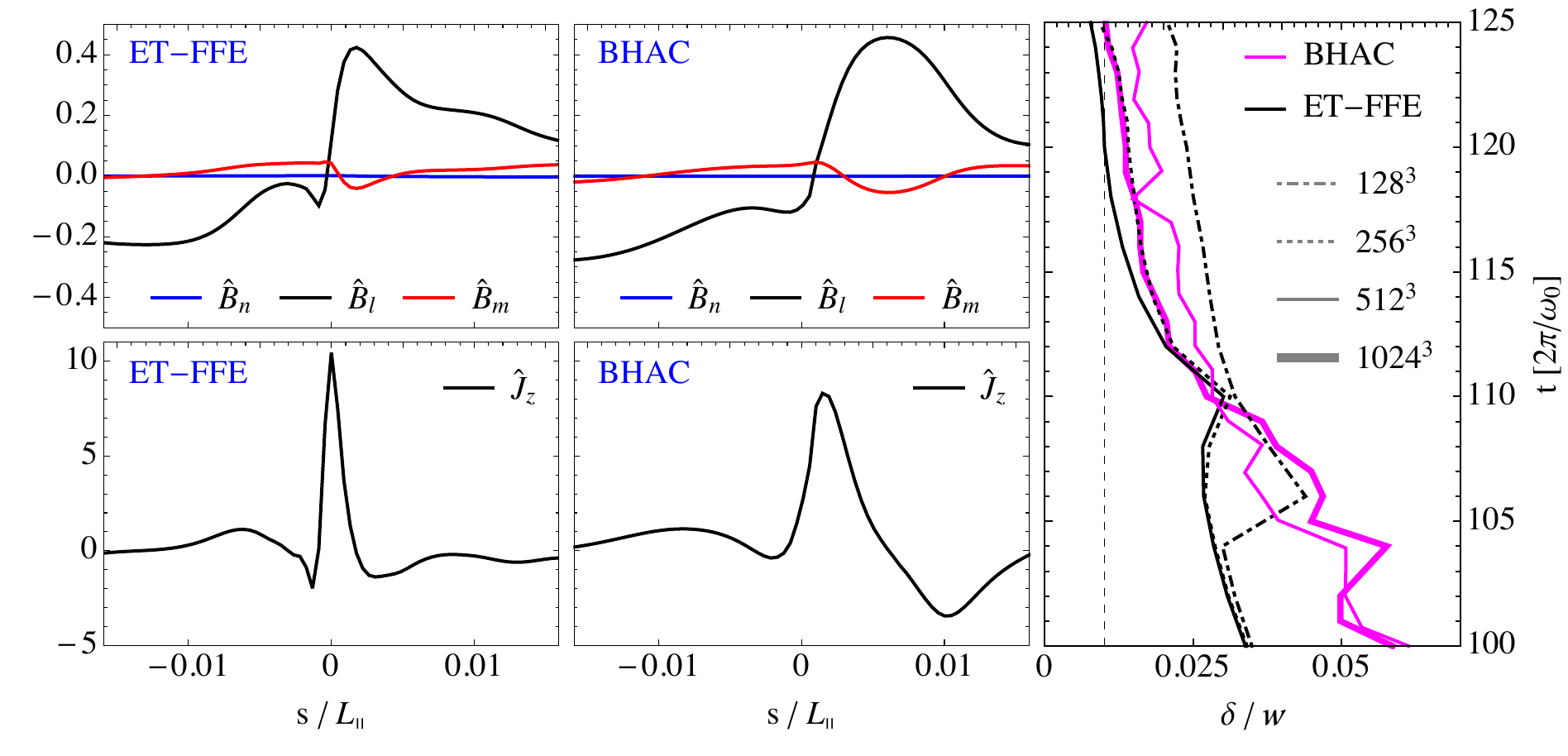}
  \vspace*{-24pt} 
    \begin{flushleft}
    \textcolor{blue}{\texttt{BHAC}/\texttt{ET-FFE}}
    \end{flushleft}
    \caption{{Current sheets form similarly across different codes and resolutions.} Profiles of normalized magnetic field $\hat{B}_i = \delta B_i/B_0$ (\textit{left top} panels) in minimum variance coordinates \citep[cf.][]{Sonnerup:1967,Howes_2016}, and out-of-plane current density $\hat{J}_z = [\nabla \times \mathbf{B}]_z / (k_{\perp}B_0)$ (\textit{left bottom} panels) at $t=125\;[2\pi/\omega_0]$. The 1D outlines are extracted across the current sheet (see dashed black lines in Fig.~\ref{fig:Compression}, where the coordinate $s$ is centered at the respective structure. The \textit{right} panel shows the current sheet's aspect ratio $\delta/w$ measured as the full-width half maximum of the current density normalized by its approximately constant width $w\approx 2\pi$ versus wave-crossings for both {\tt BHAC} and {\tt ET-FFE}. The current sheet forms around the nonlinear time $t=100\;[2\pi/\omega_0]$, and the thinning process halts at around 125 wave-crossings at $\delta/w\approx0.01$, suggesting that the sheet breaks up once an asymptotic reconnection rate $v_{\rm rec}/c \sim \delta/w \approx 0.01$ (indicated by a {vertical} dashed line) is reached.}
\label{fig:CSCompare}
\end{figure}

{In this section, we analyze the formation of the sheets, their thickness, and the dynamics until the moment they break-up and magnetic energy is quickly dissipated.} As shown in Paper I, the energy transfer to smaller scales is mediated by self-consistently generated $k_\parallel=0$ modes in the $\chi \ll 1$ limit, specifically, at least until the nonlinear time of $\sim \chi^{-2} = 100$ wave-crossings. If the nonlinearly generated Alfv\'{e}n waves increase in power they can generate coherent current sheet-like structures due to constructive interference with the primary Alfv\'{e}n modes. {Although in the ideal limit, current sheets are Dirac delta-functions, at finite dissipation they can be represented by many fewer modes such as those in \citealt{Howes_2016}, who showed that the formation of
current sheets results from interference between
just five complex Fourier modes.} To study whether current sheets can form and become thin enough to provide an efficient dissipation channel, we evolve the system for 200 wave-crossings, beyond the nonlinear time of $\sim \chi^{-2} = 100$ wave-crossings. 

\begin{figure}
  \centering
  \includegraphics[width=1.0\textwidth]{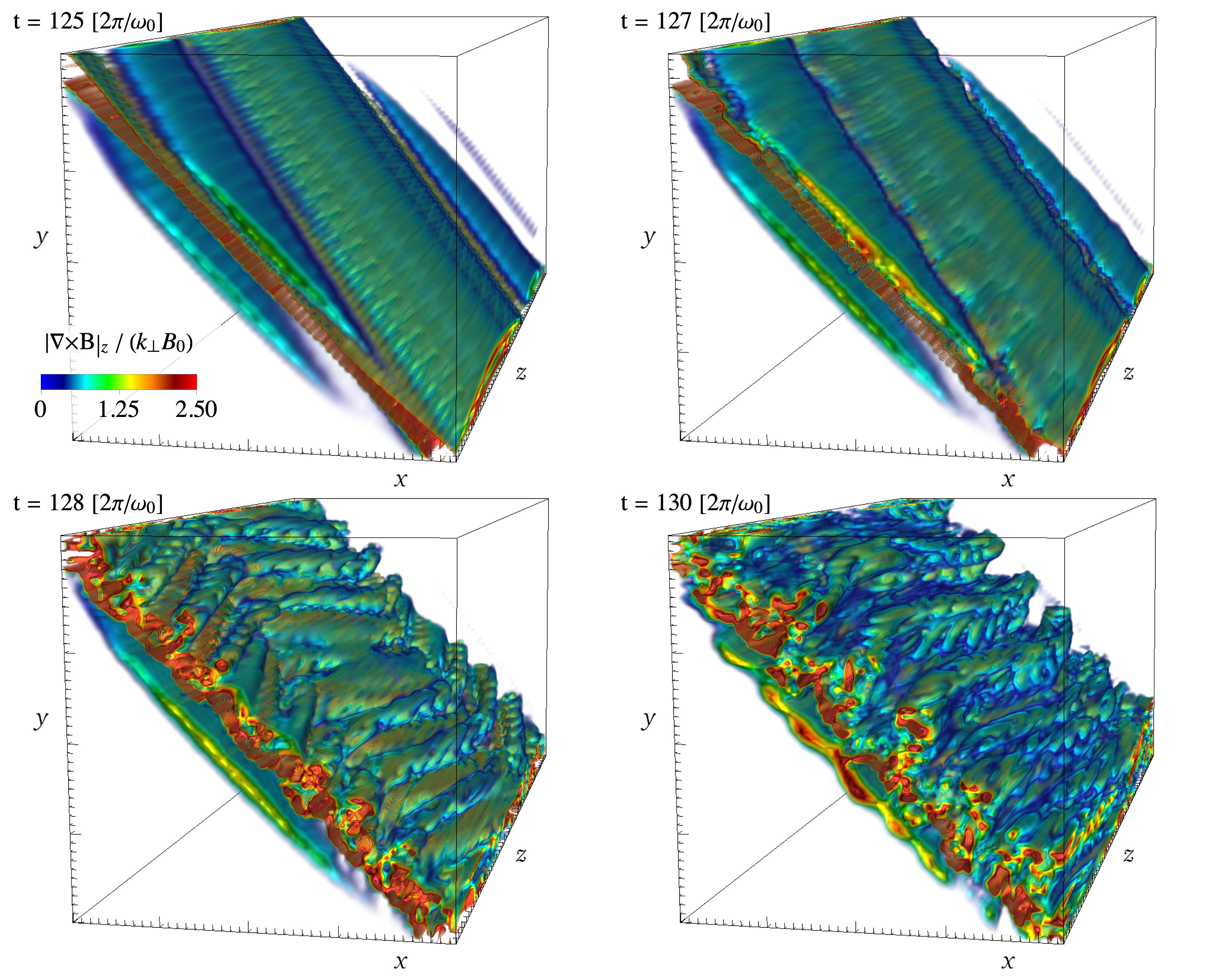}
  \vspace*{-22pt} 
    \begin{flushright}
    \textcolor{blue}{\texttt{ET-FFE}}
    \end{flushright}
  \vspace*{11pt} 
  \caption{Volume rendering of the normalized {out-of-plane current density $\hat{J}_z = [\nabla \times \mathbf{B}]_z / (k_{\perp}B_0)$}, showing the moment of the break-up of the current sheet between ${t=125\;[2\pi/\omega_0]}$ and ${t=130\;[2\pi/\omega_0]}$, within the reduced volume $x\times y\times z=\left[2\pi/5,6\pi/5\right]\times\left[2\pi/5,6\pi/5\right]\times\left[0,2\pi\right]$ at a resolution of $512^3$. {An animated version of this figure is provided in \citet{SupplementaryMediaB}.}}
\label{fig:currentsheetFF512volume}
\end{figure}

{Current sheets emerge once the wave interaction reaches a nonlinear state.} To track the dynamics of an emerging current sheet, we show the evolution of the normalized in-plane current density $\hat{J}_z = [\nabla \times \mathbf{B}]_z / (k_{\parallel}B_0)$\footnote{We confirmed that the contribution of the displacement current $d \mathbf{E} / dt$ to the current density is negligible and $\mathbf{J} \approx \nabla \times \mathbf{B}$.} perpendicular to the equilibrium field $\mathbf{B} = B_0\mathbf{\hat{z}}$ at $z=\pi$ between 100 and 150 wave-crossings in Fig. \ref{fig:currentsheetFF1024}, for the {\tt ET-FFE} run with $512^3$ grid-points. {We find that the results from {\tt ET-FFE} ($512^3$) are comparable to the high-resolution runs conducted with {\tt BHAC} ($1024^3$). The high-order reconstruction methods employed in {\tt ET-FFE} capture the smallest scales with very good accuracy \citep[see also Section \ref{sec:codecomparison}, and][]{mahlmann2020computational}.} Current sheets are characterized by in-plane magnetic field line reversals (overplotted as black lines) and are clearly distinguishable by a strong current density (red and blue colors).
At the interface of interacting eddies, current sheets are compressed into thin layers{, as we examine in detail in Fig.~\ref{fig:Compression}. In Fig.~\ref{fig:currentsheetFF1024}, such regions are} indicated by anti-parallel in-plane magnetic field, for example at $x=0.9\pi, y=0.9\pi$ at $t=125\;[2\pi/\omega_0]$. We also detect regions of non-zero current density which are not associated with current sheets due to the lack of anti-parallel magnetic field, i.e., there is no in-plane magnetic null, as in Fig.~\ref{fig:currentsheetFF1024} at $x=0.7\pi, y=0.7\pi$ at $t=110\;[2\pi/\omega_0]$.

The current sheets in this system can be characterized by the aspect ratio in the perpendicular plane between their thickness, $\delta$, and width, $w$. We determine $\delta$ as the full-width half maximum of the out-of-plane current density $\hat{J}_z$. The {sheet becomes} thinner between 100 and 127 crossings, until {it} breaks up. The width of the sheet $w$, namely, its extension along the interface of two eddies, is determined by the perpendicular wavelength $\lambda_{\perp}=2\pi$ of the initial waves in the box. It remains approximately constant $w\approx 2\pi$ from its formation at 100 crossings until the sheet breaks up at 127 crossings {(see also Fig.~\ref{fig:CSCompare})}. {The aspect ratio decreases} from {$\delta/w \gtrsim 0.025$ ({\tt ET-FFE}) at} 100 wave-crossings onward to $\delta/w \approx 0.01$ {({\tt ET-FFE}/{\tt BHAC})} at 125 wave-crossings (see also the bottom-left panels of Fig. \ref{fig:CSCompare}), indicated by the black dashed line). {The reconnection rate is expected to scale with the aspect ratio, $v_{\rm rec}/c \sim \delta/w$, suggesting that the sheet breaks up in the linear regime of the instability (\citealt{Biskamp2000,Ni2010}), and in the nonlinear regime, the asymptotic  ``fast reconnection rate'' \citep{bhattacharjee2009,uzdensky2010,Huang2016} of $0.01$ might be reached.} The life-time of the sheets is approximately 25 wave-crossings, {hence, it is maintained on time scales (and length scales) of the initial large-scale waves \citep[cf.][]{Howes_2018}}.
Turbulent fluctuations initiate inside the current sheets after 125 wave-crossings, where energy is quickly dissipated {at} the grid scale, which is smaller than the current sheet thickness, $\Delta x / \delta \lesssim 0.2$.


The 3D structure of the current sheet is lucidly illustrated in Fig. \ref{fig:currentsheetFF512volume}. We provide a volume rendering of $\hat{J}_z$ 
and observe a clear coherent structure that breaks after 127 wave-crossings. Initially, the current structure is, indeed, sheet-like in 3D. In the direction parallel to the guide field, the characteristic scale is the current sheet's length, $L$. We measure the length of the sheet along the $z$-axis to be $L\approx 2\pi$, which is determined by the parallel wavelength $\lambda_{\parallel}=2\pi$ along the equilibrium magnetic field. At the same time, it becomes visually evident that the extension in the plane perpendicular to the guide field is less than the box diagonal, but rather corresponds to $w\approx 2\pi$ (see above). At $t=125\;[2\pi/\omega_0]$, the current sheet is already very thin (see also Fig.~\ref{fig:currentsheetFF1024}). In the very short time interval of ${\Delta T}\approx 5\;[2\pi/\omega_0]$, the current sheet ripples and breaks up into turbulent structures in the plane perpendicular to the guide field. During this time, we identified that non-ideal electric fields build up in the localized region of the current sheet. Such fields are subsequently damped to maintain a (globally) force-free field configuration. 

\subsection{Comparison of current sheet characteristics in {\tt ET-FFE} and {\tt BHAC}}
\label{sec:codecomparison}

{The} emerging current sheet is analyzed in detail in Figs.~\ref{fig:Compression} and~\ref{fig:CSCompare} and compared between the employed numerical schemes. We illustrate the current sheet properties and their numerical resolution in Fig.~\ref{fig:Compression}. There, we plot numerical blocks of $8\times8$ cells in a zoom-in on the strongest currents in the domain, showing that the current sheet is indeed resolved by more than one block (approximately 11 cells) at the break-up time when the sheet is at its thinnest point, in the case of $\texttt{BHAC}$ with a total of $1024^3$ cells, and by slightly less than one block (approximately 8 cells) in the case of $\texttt{ET-FFE}$ with a total of $512^3$ cells. {While such resolution arguments do not imply a convergence of dissipative dynamics in the force-free current sheet, they correspond to the minimum resolution needed to capture their formation in the magnetically dominated regime \citep[as was established in][namely, 5 to 10 cells per current sheet width]{mahlmann2020computational}.} A notable characteristic of the examined sheet can be drawn from the overplotted contours of the out-of-plane vector potential $A_z$ (green lines): one can see that the regions of high current density originate from magnetic field compression at the edges of the elongated eddies.

{The current sheet formation and evolution is very similar for the different codes and resolutions.} In the left panels of Fig.~\ref{fig:CSCompare}, we show the normalized magnetic field $\hat{B}_i = \delta B_i/B_0$ (top) and the normalized out-of-plane current density $\hat{J}_z = [\nabla \times \mathbf{B}]_z / (k_{\parallel}B_0)$ (middle) along a cut perpendicular to {the} current sheet {presented in Fig.~\ref{fig:Compression}} just before it breaks at the time $t = 125\;[2\pi/\omega_0]$. In this analysis, we choose minimum variance coordinates \citep[cf.][]{Sonnerup:1967,Howes_2016} for comparability. These coordinates combine the guide field direction $\mathbf{\hat{m}}=\mathbf{\hat{z}}$, as well as the in-plane directions $\mathbf{\hat{n}}=\left(\mathbf{\hat{x}}+\mathbf{\hat{y}}\right)/\sqrt{2}$ and $\mathbf{\hat{l}}=\left(\mathbf{\hat{x}}-\mathbf{\hat{y}}\right)/\sqrt{2}$. For both \texttt{BHAC} and \texttt{ET-FFE}, we use the location of the strongest current in the domain for the analysis (indicated by the black dashed line in Fig.~\ref{fig:Compression}). The current sheet shows a typical profile with a peak in the current density and magnetic field reversals of the reconnecting in-plane components $\hat{B}_x$ and $\hat{B}_y$ (i.e. the perpendicular component $\hat{B}_l$). The gradient of the guide field pressure (red line) is non-zero in the current sheet, compensating for gas pressure that is absent in the force-free limit of MHD, {and} effectively {working to sustain} the evolving force-free current sheet by a {locally balanced} magnetic pressure {at the location of the field reversal} \citep[similar to the stationary force-free current sheets analyzed by][]{Komissarov_2007,Del_Zanna_2016,mahlmann2020computational}.

{The process of current sheet formation and thinning is independent of the manner in which force-free violations are managed (see also Appendix B) and is likely dominated by the field compression driven by mode interactions.} The thickness $\delta = 0.01w$ after 125 crossings is approximately 10 times larger than the smallest resolvable perpendicular scale in the {\tt BHAC} simulation, $\sim 2\pi/1024$ and {five} times larger than the smallest resolvable scale in the {\tt ET-FFE} simulation, $\sim 2\pi/512$ {(right panel of Fig.~\ref{fig:CSCompare}). Additionally, we measure the thickness of the current sheet in time for progressively increasing resolutions in the right panel, showing that the thinning rate and aspect ratio in {\tt ET-FFE} converges for $256^3$ and $512^3$ grid points, while for {\tt BHAC} it converges for $512^3$ and $1024^3$ grid points. The thinning rate and aspect ratio also converges between the highest resolution results of the two different algorithms.} We {demonstrate} that the current sheet's thickness is larger than a single numerical cell by measuring the number of cells across the sheet's thickness at 125 wave-crossings just before the sheet breaks up. The observation that the current sheet formation, thinning, as well as asymptotic thickness and time of break-up {coincide} between runs in {\tt BHAC} and {\tt ET-FFE} {(with a largely different order of convergence and treatment of the force-free violations)} is worth mentioning in this context. Small local differences aside, the comparison of results from the two numerical schemes in Figs.~\ref{fig:Compression} and~\ref{fig:CSCompare} shows that the field structure is remarkably similar in both cases. The global dynamics of field compressions is captured accurately in the force-free limit of MHD. In Appendix B, we compare the appearance of non-ideal electric fields in both methods, concluding that despite numerical differences, the results show a remarkable level of similarity both in magnitude and location of the force-free violations.


In summary, the following features let us conclude that the identified structures are 3D current sheets: a) strong and localized ordered current structures; b) field reversals with a compensating magnetic pressure; c) compression and subsequent thinning of the current structure; and, d) non-ideal electric fields around magnetic null-lines (see Appendix B) in the plane perpendicular to the guide field.

\subsection{Weak turbulence development}
\label{sec:weakturbulence}
\begin{figure}
  \centering
  \includegraphics[width=0.95\textwidth]{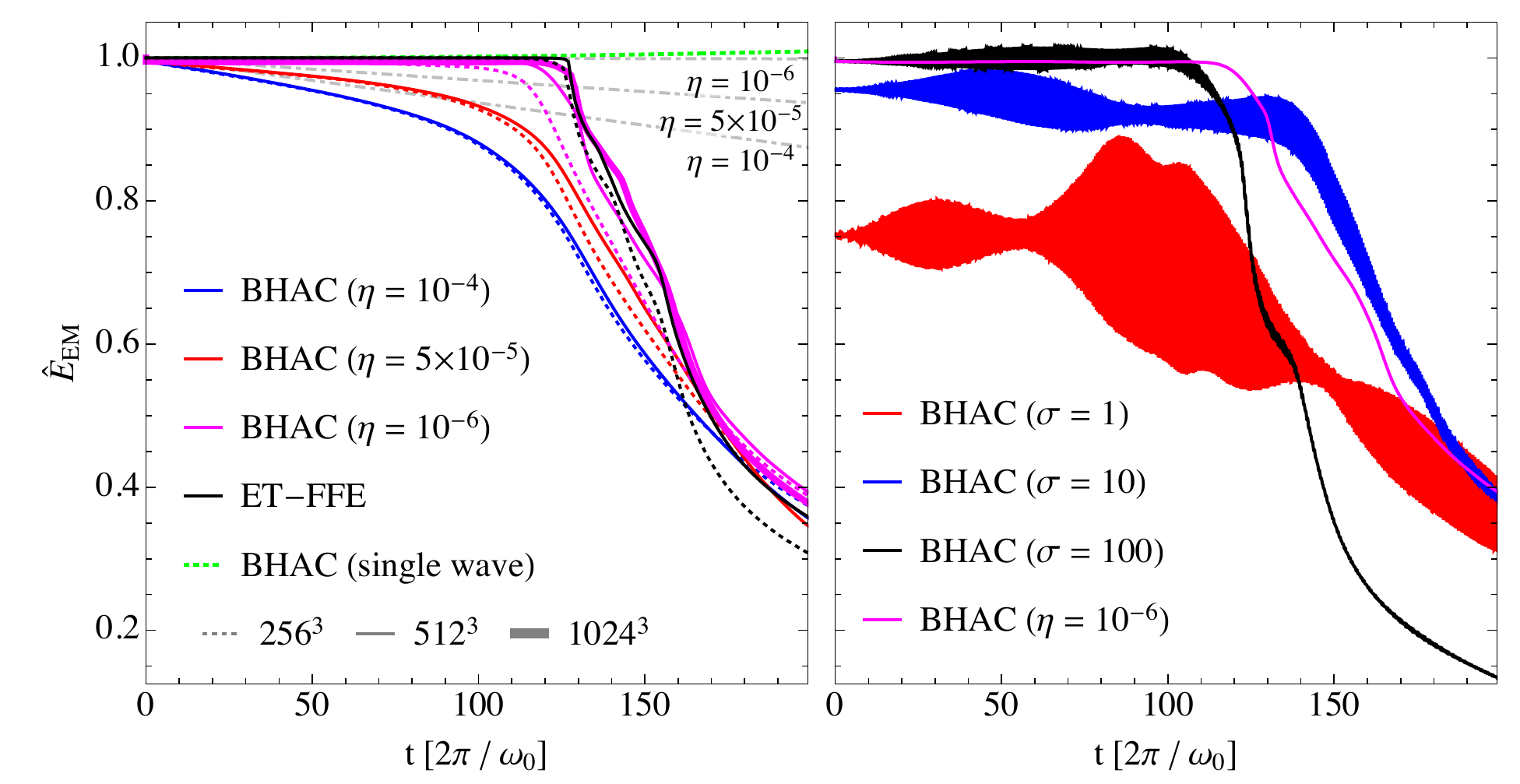}
  \vspace*{-9pt} 
    \begin{flushright}
    \textcolor{blue}{\texttt{BHAC}/\texttt{ET-FFE}}
    \end{flushright}
  \caption{Evolution of the {(normalized) total wave energy $\hat{E}_{\rm EM} = E_{\rm EM} /  E_{\rm EM,t=0}$, with $E_{\rm EM}=(B-B_0)^2 + E^2$} for the overlapping Alfv\'{e}n waves in all force-free MHD cases {at different resolutions} (\textit{left}){. We present} ideal MHD cases {and a force-free reference case }(\textit{right}{, {\tt BHAC}}) {normalized to the initial electromagnetic} wave energy of the force-free setup for 200 wave-crossings, with initial wave-amplitude $\chi=10^{-1}$ and $N_{\perp}=N_{\parallel} = 512$ grid-cells. In the left panel, we overplot results for {\tt BHAC} {(blue, red, and magenta lines for different resistivities; dotted, solid and thick lines for different resolutions)} at resolutions $N_{\perp}=N_{\parallel} = 256${, $512$} and $1024$ {as well as} {\tt ET-FFE} {(black lines)} at $256$ and $512$ grid cells to show that the energy evolution and current sheet formation converges. We also {show} the (conserved) energy density for a single Alfv\'{e}n wave at 256 cells/wavelength {(green dotted line)}, which does not undergo a cascade (\textit{left}). {The theoretical expectation of the energy dissipation for a fixed resistivity is indicated by gray dash-dotted lines, with the respective value of $\eta$ shown below the respective line.}}
\label{fig:energyxi01}
\end{figure}

In this section, we investigate how energy cascades and dissipates at small scales. In force-free MHD, the total electromagnetic energy is conserved in the domain until the turbulent cascade reaches the numerical grid scale. By measuring the total electromagnetic wave-energy $E_{\rm EM} = (B-B_0)^2+E^2$ we determine the dissipation rate depending on the resolution and the explicit resistivity{, $\eta$} (in {\tt BHAC} runs). For a resolved simulation, i.e., where the current sheet thickness is resolved by multiple cells, the energy $E_{\rm EM}$ should dissipate as $dE_{\rm EM}/dt \sim \mathbf{E} \cdot \mathbf{J} \sim \eta J^2 \sim \eta |\nabla \times \mathbf{B}|^2$, where $\nabla \times \mathbf{B} \sim B_k/\lambda$ {and $\lambda=2\pi$ is the wavelength}, implying that the typical dissipation time is $\tau \sim \lambda^2\eta^{-1}$, such that we can approximate the total energy as $E_{\rm EM}(t) \approx E_{\rm EM,t=0}(1-\eta \frac{2\pi t}{\omega_0})$. In the left panel of Fig. \ref{fig:energyxi01}, we show the estimated dissipation rate (dash-dotted lines) for force-free MHD runs with $\eta=10^{-4}$, ${5\times}10^{-5}$, and $10^{-6}$. We conclude that for $\eta\leq10^{-6}$ and resolutions of $\gtrsim 512^3$, Ohmic dissipation has a negligible effect on the initial 130 wave-crossings before the current sheets break-up. The energy dissipation rate agrees very well between {\tt ET-FFE} {($512^3$ cells)} and {\tt BHAC} {($\eta=10^{-6}$)} for a resolution of {$1024^3$} cells. This result demonstrates that the formation and the thinning process of the current sheets, occurring between 100 and 125 wave-crossings, does neither result in a significant energy decay nor in a violation of the ideal force-free constraints that would reduce electromagnetic energy density (see also Appendix B). After approximately 125 wave-crossings, a steep decay is observed, which corresponds to the break-up of current sheets in Fig. \ref{fig:currentsheetFF1024}, and energy is quickly dissipated at the grid-scale. We {further conclude} that the energy decay is a result of the cascade initiated by the wave interactions by showing that a single non-interacting Alfv\'{e}n wave {(green dashed line in the left panel of Fig.~\ref{fig:energyxi01})} shows no significant energy loss over 200 wave-crossings as indicated in the left panel of Fig. \ref{fig:energyxi01}. 
In the right panel we show the MHD runs {({\tt BHAC})} at $\sigma=1$, $10$, and $100$, corresponding to smaller initial total electromagnetic energy, showing a similar evolution with a larger variability due to the transfer of electromagnetic energy to fluid (kinetic and thermal) energy.

\begin{figure}
  \centering
  \includegraphics[width=0.95\textwidth]{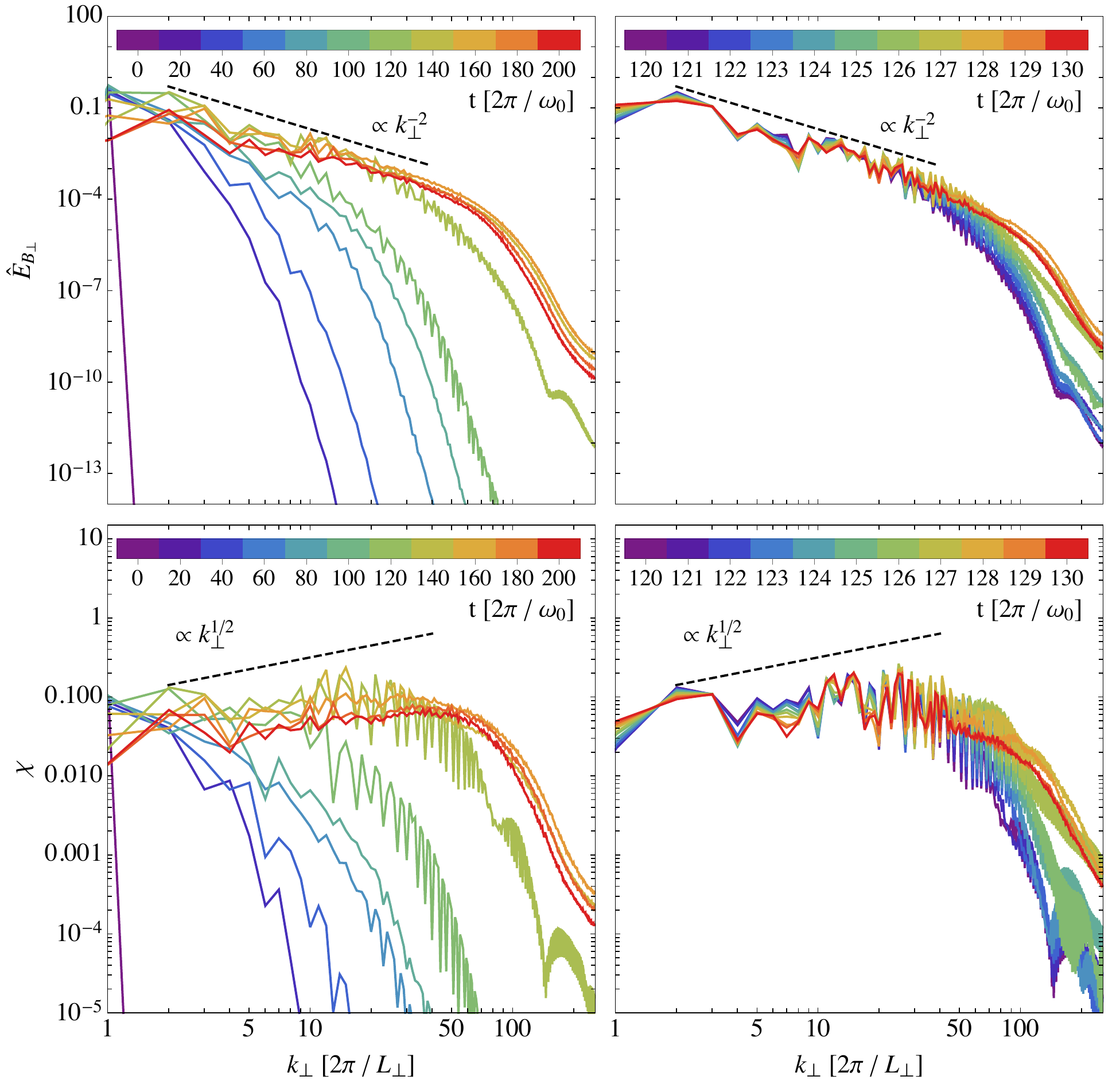}
  \vspace*{-9pt} 
    \begin{flushright}
    \textcolor{blue}{\texttt{ET-FFE}}
    \end{flushright}
  \caption{{Evolution of spectra (top row) and non-linearity parameter $\chi$ (bottom row) during the continuous wave interaction in the force-free limit. Top l}eft: $\hat{E}_{B_\perp}$ spectra at every 20 wave-crossings with wave-amplitude $\chi=10^{-1}$, and resolution $N_{\perp}=N_{\parallel} = 512$ grid-cells until 200 wave-crossings in {\tt ET-FFE}. 
  {Top r}ight: spectra at every wave-crossing between 120 and 130 wave-crossings. {Bottom row: Same analysis in time for the non-linearity parameter $\chi$ (reduced to its maximum value along slices of equal $k_\parallel$).} There is a clear transition between 120 and 130 wave-crossings, when the current sheets thin and break-up after 127 wave-crossings, and an extended $\hat{E}_{B_\perp} \propto k_{\perp}^{-2}$ spectrum develops. We note that the spectral evolution in the {\tt ET-FFE} run at $512^3$ resolution is similar to {\tt BHAC} at $1024^3$ resolution.}
\label{fig:spectraBperpxi01ET}
\end{figure}

We analyze the evolution of the spectral magnetic energy $E_B(k)\text{d}k=\sum_{\mathbf{k}\in\text{d}k}\mathbf{B}_\mathbf{k}\cdot\mathbf{B}^*_\mathbf{k}$ for the resolved\footnote{In Appendix A, we determine the dispersion error for a single wave in a periodic 3D box for a range of resolutions, to verify that the spectra we obtained are reliable after many wave-crossings, when the original wave has decayed. This verification is necessary to avoid that, at the moment of current sheet formation, the low-$k$ modes are lost due to dispersion errors and the plasma is not in the weak turbulence regime anymore. We confirm in Appendix A that energy and dispersion errors are negligible for a single wave that is evolved for 200 wave-crossings and resolved by a few cells.} {\tt ET-FFE} run at a resolution of $512^3$ cells in Fig. \ref{fig:spectraBperpxi01ET}. We observe a magnetic energy spectrum with a $E_{B_\perp} (k_{\perp})\propto k_{\perp}^{-2}$ power law forming after 100 wave-crossings (green line in the left panel) as a result of three-wave interactions. The power spectrum indicates that the turbulence in our simulations remains in the the weak regime $k_\parallel B_0 \gg k_\perp \delta B$ \citep{Ng1997,Bhattacharjee_2001,Kuznetsov_2001,Tenbarge2021}. The transition from weak to strong turbulence is expected to occur after at least $\sim \chi^{-2}=100$ wave-crossings. When the critical balance condition $k_{\parallel} B_0 \sim k_\perp \delta B$ becomes satisfied, such that the turbulence is in the strong regime, a $E_{B_\perp} (k_{\perp})\propto k_{\perp}^{-3/2}$ energy spectrum is expected to form \citep{perez2008,Schekochihin2012,Verdini2012}. Due to limited resolution and, hence, a limited inertial range, the wave energy starts to dissipate at the grid scale after $\sim 120$ wave-crossings in our simulations such that the transition to strong turbulence is not captured here. 

When the energy cascades to the highest $k$ (or thinnest structures) after 100 wave-crossings, pronounced thin current sheets of high-amplitude current density form (see Fig. \ref{fig:currentsheetFF1024}). 
Following \cite{Howes_2016}, the emerging current sheets can be described by as few as the 5 lowest order modes. Only when smaller scale structures form in the thinning and breaking of current sheets, energy is transferred to higher order modes. This effect can be observed in the shift in the spectral evolution between 100 wave-crossings and the break-up point at 127 wave-crossings (see the right-hand panel of Fig. \ref{fig:spectraBperpxi01ET} for a zoom into the interval between 120 and 130 wave-crossings). At this point, the inertial range extends for almost two decades of $k_{\perp}$. {Note that a current sheet itself results in a spectral index of $-2$ by assuming the Fourier transform of a (near)-step function (\citealt{burgers1948}).}
When the current sheets break-up at 127 wave-crossings, a maximally-developed spectrum has formed as any thinner structures cannot exist on the numerical grid.
Once the sheets break-up, the smallest scale (highest $k$) is reached by the cascade, and the magnetic energy starts to dissipate at the grid-scale {(roughly corresponding to $k_\perp>100$)}. The spectral energy decays from here onward ({orange} and {red} lines).

{The growth of turbulence at smaller scales and the subsequent decay is illustrated in the analysis of the $\chi$ parameter in the bottom rows of Fig.~\ref{fig:spectraBperpxi01ET}. The steady-state expectation of $\chi(k_{\perp})$ can be derived by assuming that the parallel cascade is negligible, and the weak turbulence spectrum shows that $\delta B_{\perp}^2 / k_{\perp} \sim k_{\perp}^{-2}$, hence $\chi = k_{\perp} v_{\perp} / (k_{\parallel} v_A) \sim \delta B_{\perp} k_{\perp} / (k_{\parallel} B_0) \sim k_{\perp}^{1/2} / k_{\parallel} \sim k_{\perp}^{1/2}$, indicated by the dashed line. The nonlinearity parameter clearly indicates that the turbulence remains in the weak regime $\chi < 1$ and develops towards $\chi \sim k_{\perp}^{1/2}$.} The increasing nonlinearity parameter with $k_{\perp}$ and the cascade becoming more anisotropic suggests that, for higher resolutions, the transition to strong turbulence would be inevitable (\citealt{Meyrand2016}).

\subsection{Local weak turbulence properties}
\label{sec:structurefunctions}
In this section, we investigate the local properties of the turbulence to determine the local anisotropy of the spectrum.
The Fourier analysis employed for the computation of the spectra cannot take into account the local properties of turbulence. It implicitly assumes the local mean magnetic field remains parallel to the initial field ${\bf B_0}$. However, each small scale eddy responds to the combined, local contribution of larger scale eddies. The weak turbulence limit requires many collisions of counter-propagating waves to substantially deform the initial eddy, and an Alfv\'{e}nic eddy propagates along the local, mean magnetic field. Accounting for the mismatch of \textit{local} and \textit{global} guide field requires one to define locally parallel and perpendicular cascades \citep{ChoVishniac,Maron:2001}. To mitigate the effect of mixing parallel and perpendicular directions, we employ second order structure functions 
\begin{equation}
\delta B_i^2 ({\bf l}) = \langle \left\lvert B_i({\bf r}+{\bf l})-B_i({\bf r}) \right\lvert^2\rangle,
\label{structFunction}
\end{equation}
where $l$ is a separation vector, and angular brackets denote averages over all ${\bf r}$. These functions contains the information about the local spectrum $f({\bf r_0}, {\bf l})$ at a given point ${\bf r_0}$, where $l\sim k^{-1}$. The local guide field for two points of an eddy of size $l$ can be defined as
\begin{equation}
 {\bf B}_{\rm guide}({\bf r},{\bf l}) = \frac{{\bf B}({\bf r}+{\bf l})+{\bf B}({\bf r})}{2}.
\end{equation}
The shape of an eddy is given as $l_\parallel={\bf l \cdot B}_{\rm guide}/|{\bf B}_{\rm guide}|$, $l_{\perp} = (l^2-l^2_\parallel)^{1/2}$, assuming that all eddies are isotropic in the field-perpendicular plane and neglecting effects of dynamical alignment and intermittency \citep{Boldyrev_2005, Boldyrev_2006, Chandran_2015,Mallet:2015}. The construction of (\ref{structFunction}) requires calculation of a 6D integral, and we employ a Monte-Carlo method with a total of $N=N_0 n_x n_y n_z$ randomly chosen pairs of vectors to uniformly cover both the space of positions of eddies $\bf r$ and their sizes $\bf l$. Limits for the point-separation vectors $l_i$ are chosen to be $(-n_i/2,n_i/2)$ to include the effects of the periodic boundary conditions in direction $i$. We set $N_0=30$ to decrease the shot-noise and confirm convergence of the results for $N_0=2,5,20$.


\begin{figure*}
  \centering
\includegraphics[width=0.9\linewidth]{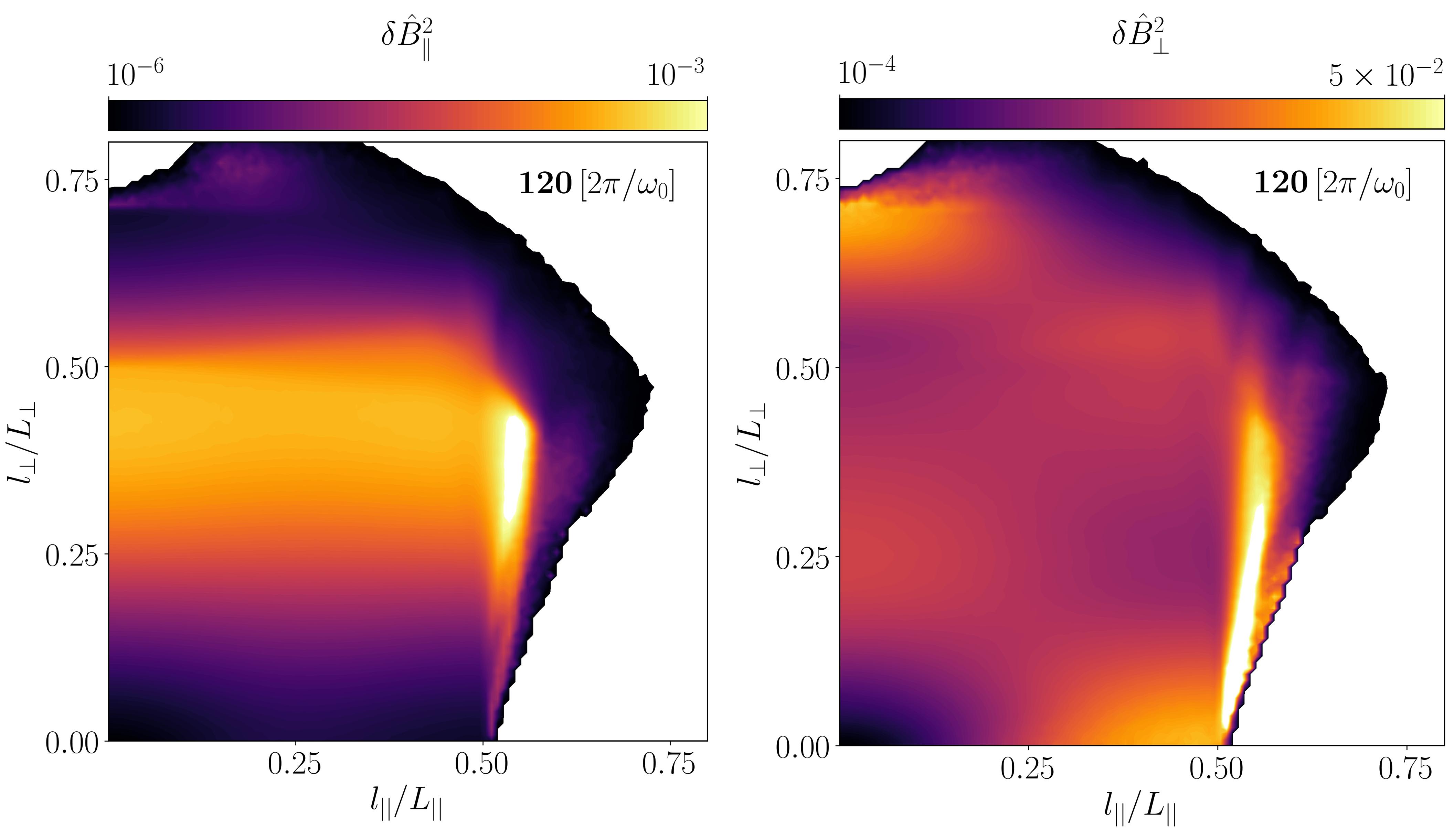}
  \vspace*{-12pt} 
    \begin{flushright}
    \textcolor{blue}{\texttt{ET-FFE}}
    \end{flushright}
\caption{2D structure functions for (left panel) {$\delta \hat{B}_\parallel$} and (right panel) {$\delta \hat{B}_{\perp}$} with respect to the local guide field, at time $t=120\;[2\pi/\omega_0]$ before the current sheets break-up. $l_\parallel$ and $l_\perp$ are defined with respect to the local guide field. White regions at large $l$ do not have any points due to periodic boundary conditions. The bright feature at $l_\parallel/L_\parallel\approx0.5$ corresponds to the current sheets.}  
\label{fig::2dStructureFunctions}
\end{figure*}

\begin{figure*}
  \centering
\includegraphics[width=1\linewidth]{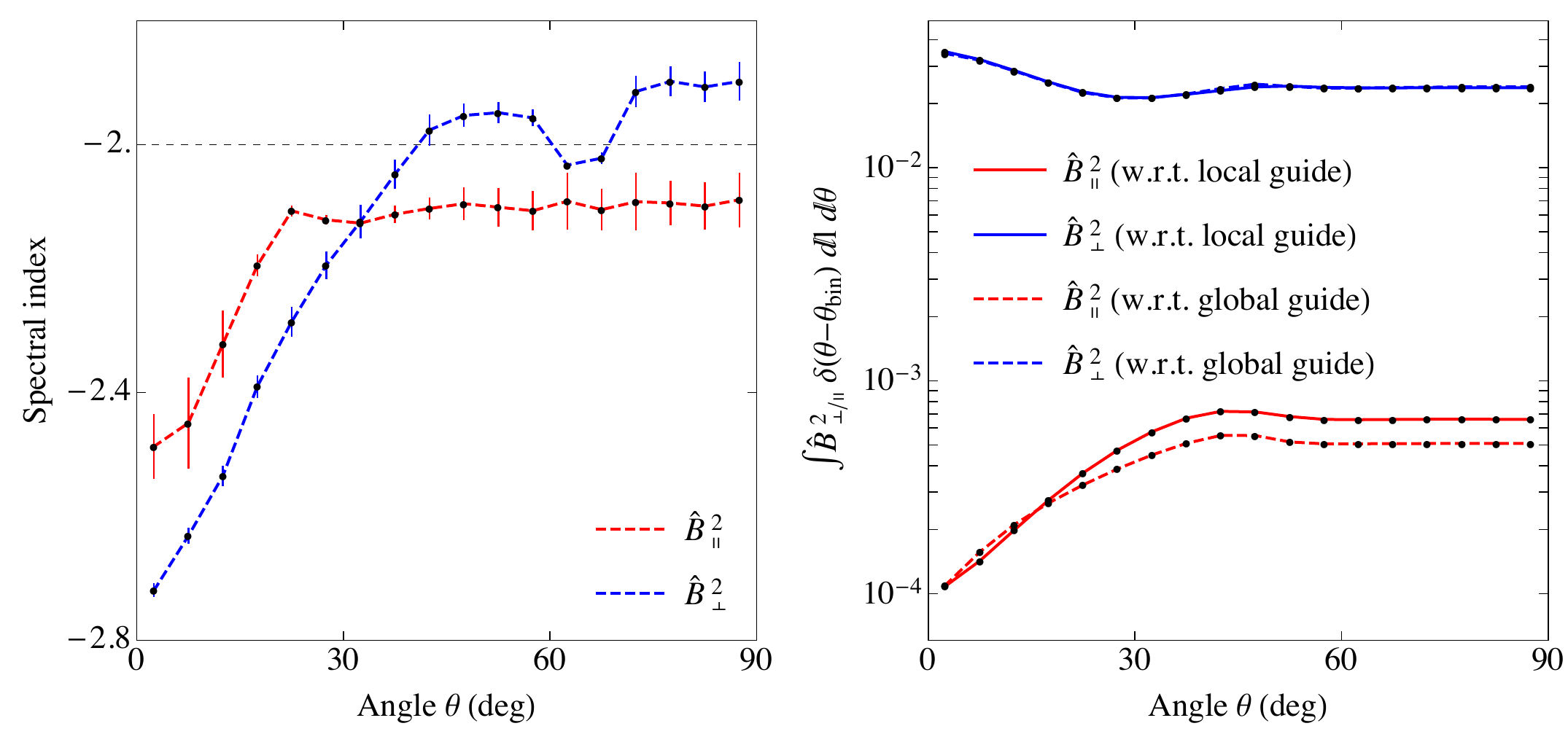}
  \vspace*{-20pt} 
    \begin{flushright}
    \textcolor{blue}{\texttt{ET-FFE}}
    \end{flushright}
\caption{Angular properties for the turbulence at $t=120\;[2\pi/\omega_0]$: the left panel shows the spectral index of the power-law spectrum as a function of angle between the $\textit{local}$ guide field and a chosen direction ($\theta$), the right panel shows the total energy contained in modes perpendicular and parallel to the $\textit{local}$ (solid lines) and $\textit{global}$  guide field (dashed lines) modes.}  
\label{fig::anisotropy}
\end{figure*}

The 2D structure functions for an {\tt ET-FFE} run at a resolution of $512^3$ cells are shown in Fig. \ref{fig::2dStructureFunctions}, at time $t=120\;[2\pi/\omega_0]$, when thin current sheets have formed and are not yet broken up (see Fig. \ref{fig:currentsheetFF1024}). All vectors ${\bf l}$ are split into $128$ bins in both parallel and perpendicular directions. The white regions at larger $l_\parallel$ and $l_{\perp}$ do not contain any points as all separations larger than $l_i/2$ are prohibited due to the periodic grid. The shape of the structure function is defined by the winding of magnetic field lines. A bright over-scaled feature at $l_\parallel \sim 0.5 L_\parallel$ and $l_\perp < 0.5 L_\perp$ is associated with the thinning current sheets: {its} amplitude grows while current sheets are being formed, and disappear when current sheets break-up at $t\approx128$. This feature appears both in the {\tt BHAC} and {\tt ET-FFE} results and is independent of how force-free violations are handled (see Appendix B).

The natural ability of this approach to consider local guide field variation allows us to measure a local angular anisotropy of the spectrum. To perform such a calculation, we split all point-separating vectors into 18 angular bins of $5^\circ$ extension, measured with respect to the field-parallel direction ($0^\circ$ is parallel to the guide field and $90^\circ$ is perpendicular to it). In each bin, the intermediate range of $|{\bf l}|$ is approximated by a power-law function $\propto l^g$. The spectral index $\alpha$ is found using the relation $\alpha= -(g+1)$ \citep{monin_yaglom_1999}. The resulting anisotropy is shown in the left panel of Fig. \ref{fig::anisotropy}, and we observe that the spectrum is steeper in the parallel direction than in the perpendicular. For $\theta \rightarrow 0$, where $\theta$ is the angle between the $\textit{local}$ guide field and a chosen direction, the spectral index steepens to $-3$, which is the maximum that can be recovered using second order structure functions as applied here \citep{Farge:2006}. This result implies that there is no parallel cascade. For $\theta \gtrsim 20$ we find that the $E_{B_\parallel}$ spectrum is isotropic, indicating the expected isotropy of a fast wave $E_{B_\parallel}$ spectrum \citep{Cho:2002a,Cho:2003,Chandran:2005}. In the right panel of Fig. \ref{fig::anisotropy}, we present the energy in modes parallel and perpendicular to the local guide field. This allows us to estimate the energy distribution in Alfv\'{e}n and fast waves. The latter have parallel perturbations of the magnetic field while Alfv\'{e}n waves have only perpendicular fluctuations $\delta {\bf B} \perp {\bf B}_{\rm guide}$. Most of the magnetic energy is associated to Alfv\'{e}n waves, and the fraction of energy in fast waves remains low at $1-5\%$. For comparison, we overplot the energy distribution in the parallel and perpendicular components to the global guide field $B_0$: they are shown with dashed lines on the right panel of Fig. \ref{fig::anisotropy}. The fact that the energies match closely with respect to local and global fields highlights the nature of weak turbulence: as $\tau_{nl} \gg \tau_A$, eddies predominantly interact with the $global$ guide field. The difference in the energy distribution of the parallel modes can be explained by projection of the energy in perpendicular modes on the parallel direction.

\subsection{Weak turbulence and current sheets in relativistic magnetohydrodynamics}
\begin{figure}
  \centering
  \includegraphics[width=0.95\textwidth]{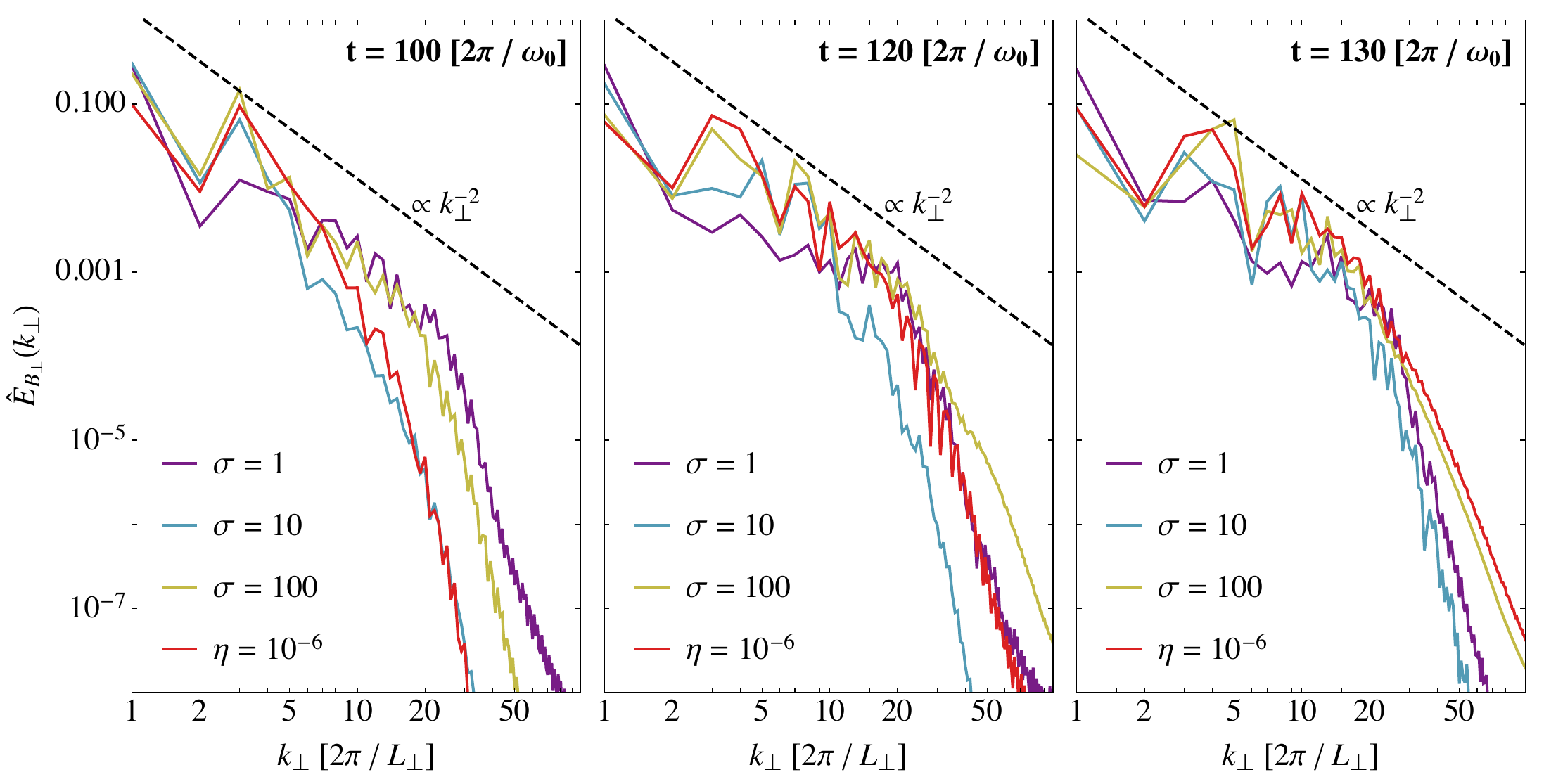}
  \vspace*{-9pt} 
    \begin{flushright}
    \textcolor{blue}{\texttt{BHAC}}
    \end{flushright}
  \caption{{Evolution of spectra during the continuous wave interaction for different magnetizations.} $\hat{E}_{B_\perp}$ spectra at $100\pm 2$, $120\pm 2$, and $130\pm 2$ wave-crossings at resolution $N_{\perp}=N_{\parallel} = 512$ grid-cells for runs at $\sigma=100$, $10$, and $1$ and force-free MHD in {\tt BHAC}. The spectra are averaged over 5 wave-crossings. For the force-free MHD run, we set $\eta=10^{-6}$. $\hat{E}_{B_\perp} \propto k_{\perp}^{-2}$ spectra develop at 100 wave-crossings.}
\label{fig:spectraBperpMHD}
\end{figure}

{In this section, we compare the formation of current sheets in weak MHD turbulence to the force-free results presented in the previous sections.}
In Fig. \ref{fig:spectraBperpMHD}, we compare force-free MHD and ideal MHD spectra obtained with {\tt BHAC} for $\sigma=1$, $10$, and $100$ at $512$ cells per initial wavelength, showing that in all four cases a $E_{B_\perp} (k_{\perp})\propto k_{\perp}^{-2}$ magnetic energy spectrum forms and then a transition occurs between 100 (left panel) and 130 (right panel) wave-crossings induced by current sheet formation and break-up. After 120 wave-crossings (middle panel) the total electromagnetic energy starts to decay (see right panel of Fig. \ref{fig:energyxi01}). Due to energy transfer between electromagnetic and hydrodynamic (kinetic and thermal) components, the onset time of the steep decay and therewith the development of the spectrum differs slightly for varying $\sigma$, where $\sigma=100$ corresponds most closely to the force-free MHD result.

\begin{figure}
  \centering
  \includegraphics[width=1\textwidth]{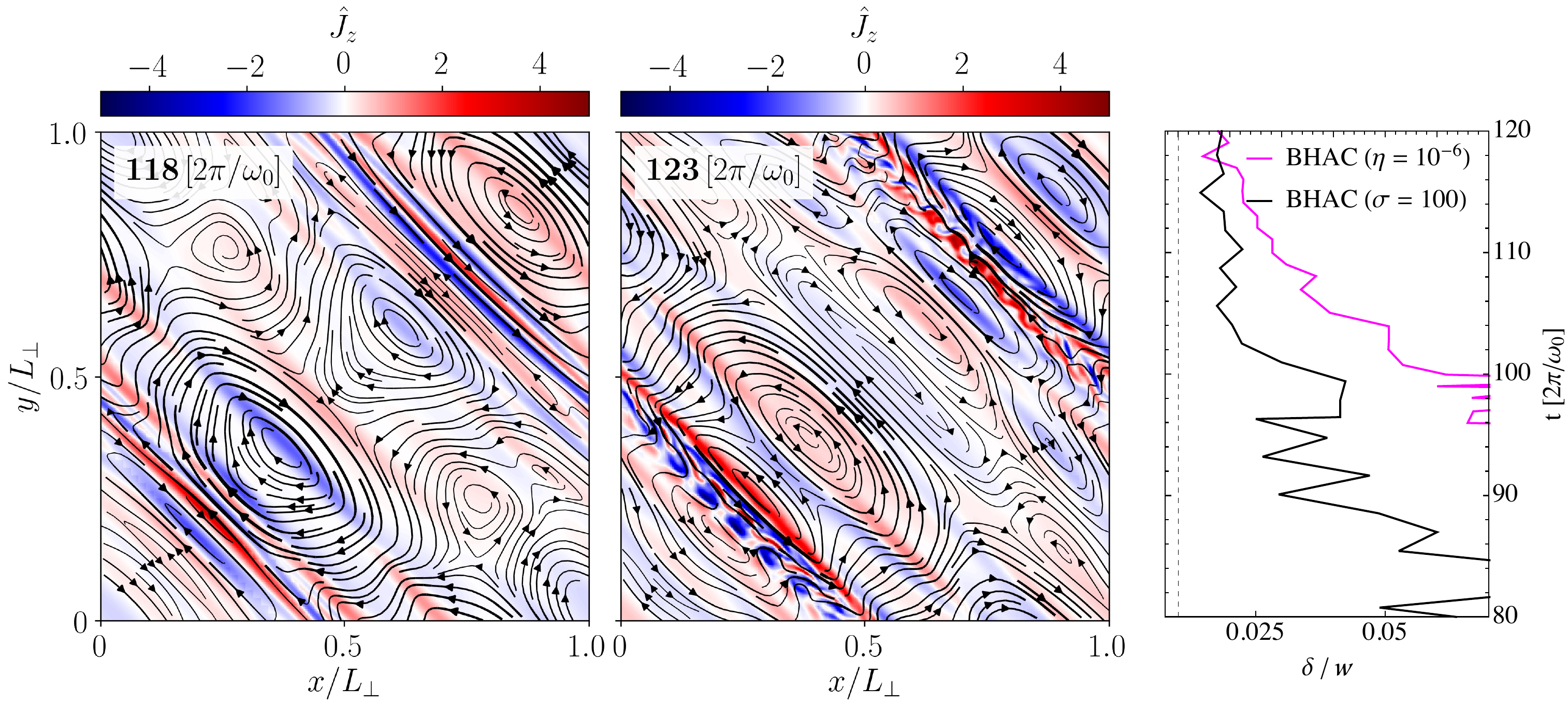}
  \vspace*{-20pt} 
    \begin{flushleft}
    \textcolor{blue}{\texttt{BHAC}}
    \end{flushleft}
    \caption{{Current sheets form between elongated eddies in the MHD run for $\sigma=100$ with {\tt BHAC}. The \textit{left} and \textit{middle} panel show the out-of-plane current density $\hat{J}_z = [\nabla \times \mathbf{B}]_z / (k_{\perp}B_0)$ at 118 and 123 wave-crossings, respectively, before and after the break-up of the current sheet. The \textit{right} panel shows the current sheet's aspect ratio $\delta/w$ measured as the full-width half maximum of the current density normalized by its approximately constant width $w\approx 2\pi$ versus wave-crossings for both the $\sigma=100$ and force-free runs in {\tt BHAC} at $512^3$ grid points.}}
\label{fig:BHACMHD}
\end{figure}
{We show the structure of the forming current sheet shortly before it breaks-up at $t = 118\;[2\pi/\omega_0]$ in the $\sigma=100$ MHD run in {\tt BHAC} in the left panel of Fig. \ref{fig:BHACMHD}, similar to the force-free current sheet in Fig. \ref{fig:currentsheetFF1024}. The sheet breaks up at $t = 123\;[2\pi/\omega_0]$. A pressure gradient contributes to the force balance to sustain current sheets in the relativistic MHD simulations. For smaller $\sigma$, the current sheets have a smaller aspect ratio, and they have an effectively lower Lundquist number due to the lower Alfv\'{e}n speed and length. The thinning rate (right panel) of the current sheet is very similar to the force-free result, which converges between $512^3$ and $1024^3$ grid points (see Fig. \ref{fig:CSCompare}).
}

\section{Collisions of Alfv\'{e}n wave packets}
\label{sec:alfvenpackets}

In this section, we illuminate the development of current sheets in a more realistic setting of collisions between initially separated Alfv\'{e}n wave packets (see e.g., \citealt{Verniero_2018,Verniero_2018b,Li_2019,li2021fast}). {Due to the localization of wave packets, we expect the character of secondary modes to differ significantly from the results presented in Section \ref{sec:alfvendynamics} (cf. \citealt{Verniero_2018}). In the following sections, we will dissect the secondary mode structure emerging during the interaction of Alfv\'{e}n wave packets in the relativistic limit. Further, we probe if the finite extension and interaction time of the wave also results in rapid dissipation of electromagnetic energy in current sheets, as we proposed in Section \ref{sec:currentsheetformation}.}

{\subsection{Packet initialization}
\label{sec:packetinit}}

To allow for a direct comparison to the results of the previous sections, we only modify the wave initialization described in Section \ref{sec:waveinitialization} in two key aspects. First, the cubic box is extended along the direction parallel to the guide field, with $L_\parallel=10\pi$. {This choice allows us to separate the waves in packets while} the choice of $k_\parallel$ re-scales the wave in the elongated domain. Second, the two Alfv\'{e}n waves are localized along the (parallel) $z$-direction by window functions with a Gaussian profile \citep[cf.][]{Verniero_2018}:

\begin{align}
    w(z)=\exp\left[-\left(\frac{z-z_0}{\Delta_z}\right)^p\right]+\exp\left[-\left(\frac{z-z_0+L_\parallel}{\Delta_z}\right)^p\right]+\exp\left[-\left(\frac{z-z_0-L_\parallel}{\Delta_z}\right)^p\right].
\end{align}

We choose a window width of $\Delta_z = L_\parallel / 4\pi$, and the power $p=2$. The two waves are localized at the window centers $z_{0-}=L_\parallel/4$, and $z_{0+}=3\times L_\parallel/4$. Initial phase shifts $\delta_-$, and  $\delta_+$ align the waves symmetrically in their respective windows. For the tests presented in this section, we fix the nonlinearity parameter to $\chi = k_{\perp} \delta B_{\perp}/ (k_{\parallel}B_0)=0.25$. Hence, though the wave amplitudes are comparable to the continuous setup (Section \ref{sec:alfvendynamics}), the nonlinear time in the localized setup is drastically reduced to $\chi^{-2}=16$ wave interactions. {This choice is beneficial for this numerical exploration. In contrast to the overlapping waves in Section \ref{sec:alfvendynamics}, the elongated domain requires each wave packet to propagate significant distances between the interaction events. Decreasing the nonlinear time, thus, allows us to reach the relevant stages of turbulence during a reasonable computational time.}

\begin{figure}
  \centering
  \includegraphics[width=0.9\textwidth]{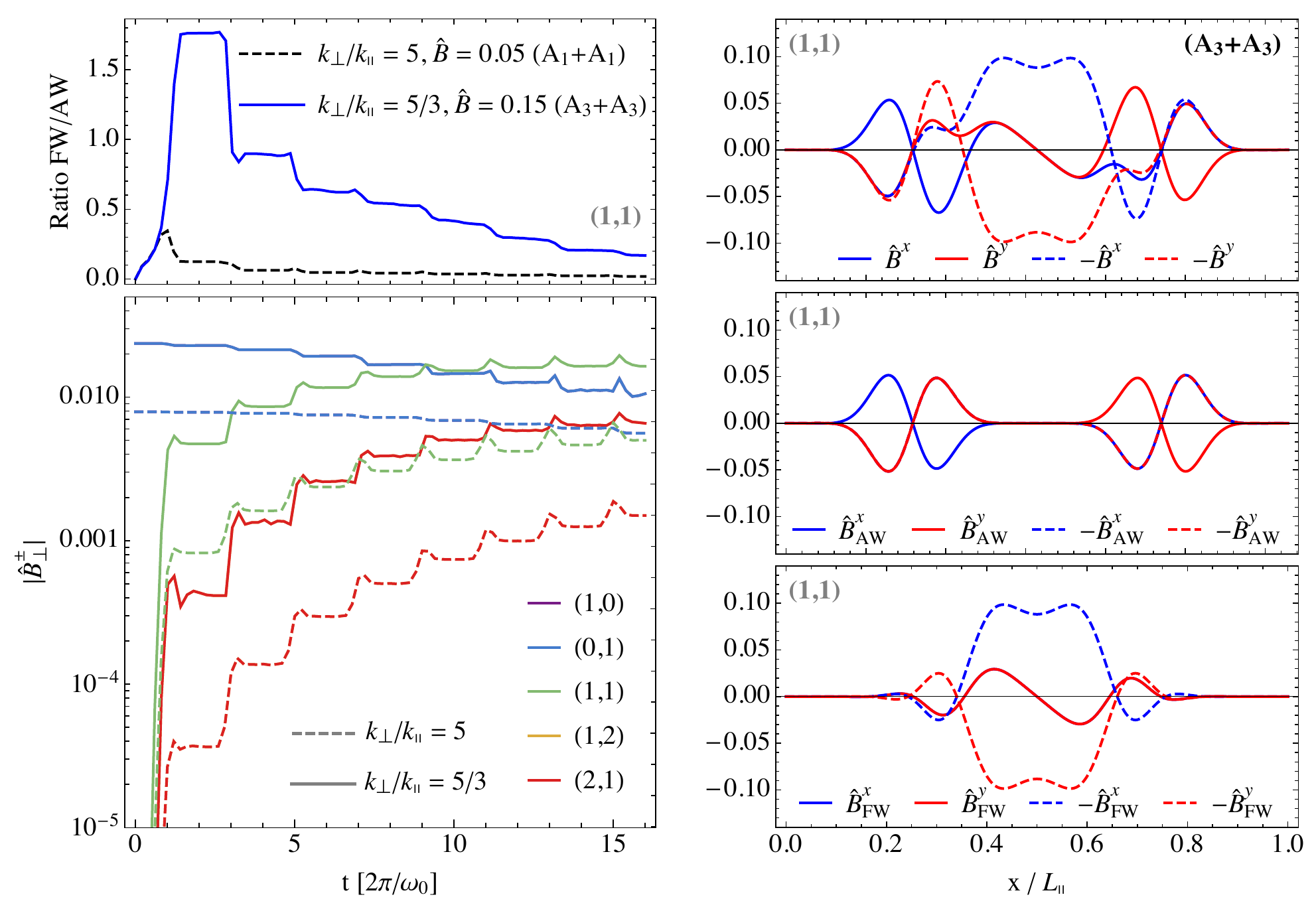}
  \vspace*{-11pt} 
    \begin{flushright}
    \textcolor{blue}{\texttt{ET-FFE}}
    \end{flushright}
  \caption{Evolution of individual modes during the interaction. \textit{Left} panel: Comparison of the ${\mathrm{AW}}_3+{\mathrm{AW}}_3$ and ${\mathrm{AW}}_1+{\mathrm{AW}}_1$ collisions {for the intermediate resolution of $256^2\times 320$}. Decay of the primary modes $(1,0)$ and $(0,1)$ with sustained growth of the $(1,1)$ mediator mode as well as the secondary modes $(2,1)$ and $(1,2)$ (\textit{bottom}). The respective ratio between the total energy stored in the Alfv\'{e}n and in the fast channel of the $(1,1)$ mode is displayed in the \textit{top} panel. \textit{Right} panel: Exemplary decomposition of the $(1,1)$ mode by its polarization for the ${\mathrm{AW}}_3+{\mathrm{AW}}_3$ collisions. We display the total magnetic field along the center of the domain (\textit{top}), the projection into the Alfv\'{e}n mode (\textit{middle}), and the projection into the fast mode (\textit{bottom}). {An animation of the (1,1) mode structure and dynamics during the first interaction cycles is provided in \citet{SupplementaryMediaC} and \citet{SupplementaryMediaH}.}}
\label{fig:modes_packetsxi025}
\end{figure}

\begin{figure}
  \centering
  \includegraphics[width=0.95\textwidth]{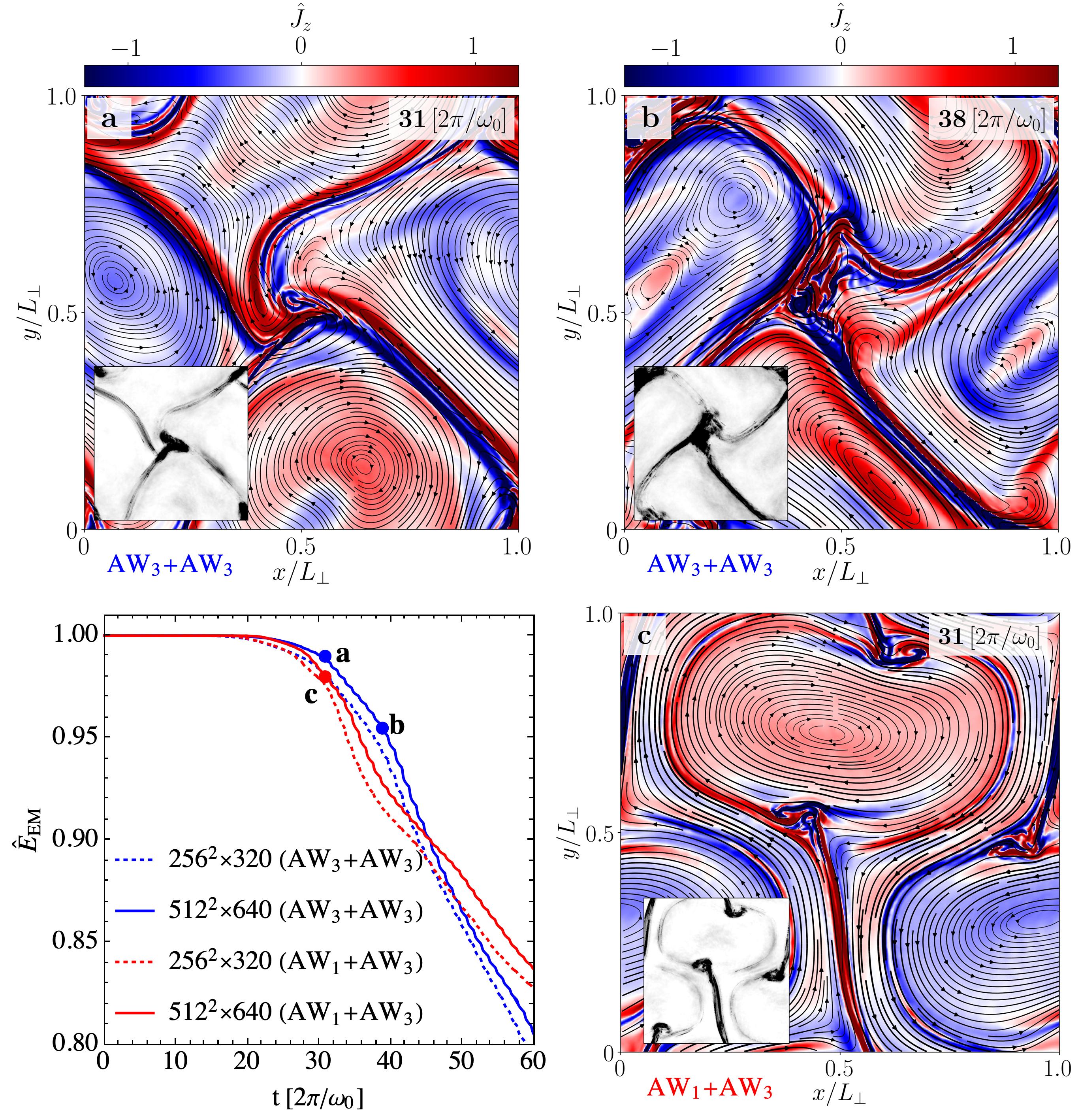}
  \vspace*{-11pt} 
    \begin{flushright}
    \textcolor{blue}{\texttt{ET-FFE}}
    \end{flushright}
  \caption{Decay of (normalized) total wave energy $\hat{E}_{\rm EM} = E_{\rm EM} /  E_{\rm EM,t=0}$, with $E_{\rm EM}=(B-B_0)^2 + E^2$ driven by the reconnection zones and current sheets induced by the shearing action of the secondary mode. \textit{Bottom left:} Evolution of the wave energy (normalized) for interactions of different wave types and resolutions. The remaining panels visualize the {out-of-plane} current {density $\hat{J}_z = [\nabla \times \mathbf{B}]_z / (k_{\perp}B_0)$} and field composition of the wave packets at different times in high resolution as indicated in the energy plot. The insets provide an impression of the $z$-averaged non-ideal field marker $\mathbf{E}\cdot\mathbf{B}/\mathbf{B}^2$. Regions of strong (numerical) diffusion of non-ideal electric fields (i.e., violations of the force-free conditions) appear in dark shades. Such locations include Y-points, such as the anchor points of the shearing motion, and current sheets. {Animations of the interaction dynamics are provided in \citet{SupplementaryMediaD} and \citet{SupplementaryMediaE}.}}
\label{fig:Packets2D}
\end{figure}

We present insight into the collisional dynamics of localized Alfv\'{e}n waves by combining two specific wave types: ${\mathrm{AW}}_1$, where $k_\parallel= 2\pi/L_\parallel$ with an initial wave amplitude of $\delta B_{\perp} / B_0=0.05$ and $k_\perp/k_\parallel=5$; and ${\mathrm{AW}}_3$, where $k_\parallel= 3\times 2\pi/L_\parallel$ with an initial wave amplitude of $\delta B_{\perp} / B_0=0.15$ and $k_\perp/k_\parallel=5/3$. {Specifically, we examine the interactions $\mathrm{AW}_1+\mathrm{AW}_1$ and $\mathrm{AW}_3+\mathrm{AW}_3$ to understand the evolution of higher order modes (Fig.~\ref{fig:modes_packetsxi025}). We then focus on comparing the interactions $\mathrm{AW}_1+\mathrm{AW}_3$ and $\mathrm{AW}_3+\mathrm{AW}_3$ by closely following the effect of shearing modes and the ensuing dissipation of electromagnetic energy (Fig.~\ref{fig:Packets2D}).} The wave vectors differ to those of the continuous case (Section \ref{sec:alfvendynamics}) in their parallel component: $k_\parallel$ is both re-scaled and filtered by the window function, resulting in the initial wave packets having a broad spectrum of $k_\parallel$ modes rather than $k_\parallel = \pm 1$. It is, thus, simplest to limit the representation of modes to the components of $k_\perp$, namely, $(k_x L_\perp,k_y L_\perp)=(1,0)$ for $\mathbf{k}^{+}_1$, and  $(0,1)$ for $\mathbf{k}^{-}_1$. The mediator mode $\mathbf{k}^{0}_2$ is then $(1,1)$, with tertiary modes $(2,1)$ for $\mathbf{k}^{+}_3$, and $(1,2)$ for $\mathbf{k}^{-}_3$. The reality condition of the Fourier transform also implies a mirrored energy in the modes $(-1,0)$ for $\mathbf{k}^{+}_1$ and $(0,-1)$ for $\mathbf{k}^{-}_1$, as well as in their respective superpositions. The interaction of Alfv\'{e}n wave packets proceeds through the coupling of these modes in (superficial) analogy to the continuous case presented in Section \ref{sec:alfvendynamics}. The details of this mechanism, however, reveal some notable subtleties.

{\subsection{Interaction dynamics}
\label{sec:interactiondynamics}}

Two \emph{localized} Alfv\'{e}n waves create linear fast waves as well as linear Alfv\'{e}n waves during their interaction, and {we exploit the mode structure in Fourier space to conduct the analysis of secondary modes. Such a transformation allows us to select specific coordinates of $k_\perp$ and $k_\parallel$ to identify and distinguish Alfv\'{e}n and fast modes.} As we show in Fig.~\ref{fig:modes_packetsxi025} (bottom left panel), the interacting primary modes $(1,0)$ and $(0,1)$ transfer their energy to the $(1,1)$ mediator mode and, subsequently, to the tertiary waves $(1,2)$ and $(2,1)$. {The energy transfer is strongest} during each collision{, as one can infer from the step-like growth in the bottom left panel of Fig.~\ref{fig:modes_packetsxi025}}. In analogy to \citet{Verniero_2018b}, we examine two key characteristics of the secondary $(1,1)$ mode: field polarization and wave dispersion. Force-free linear Alfv\'{e}n waves with dispersion $\omega(\mathbf{k})=\pm k_\parallel$ are composed of an electric field along $\mathbf{k}_\perp$ and a magnetic field along $\mathbf{\hat{z}}\times\mathbf{k}_\perp$. For the $(1,1)$ mode, this reduces to the following polarization and group velocity
\begin{align}
    [{\mathrm{AW}}]\qquad B^x = \pm E^y \qquad B^y = \mp E^x\qquad \frac{\partial\omega}{\partial k_\parallel}=\pm 1.
    \label{eq:AWpolarization}
\end{align}
Force-free fast waves with dispersion $\omega(\mathbf{k})=\pm |\mathbf{k}|$ are composed of an electric field along $\mathbf{k}_\perp\times\mathbf{\hat{z}}$ and a magnetic field along $\mathbf{k}\times(\mathbf{k}_\perp\times\mathbf{\hat{z}})$. For the $(1,1)$ mode, this reduces to the following polarization and group velocity
\begin{align}
    [{\mathrm{FW}}]\qquad B^x = \mp\frac{k_\perp}{|\mathbf{k}|} E^y \qquad B^y = \pm \frac{k_\perp}{|\mathbf{k}|}E^x\qquad \frac{\partial\omega}{\partial k}=\pm 1.
    \label{eq:FWpolarization}
\end{align}

{Both Alfv\'{e}n waves and fast waves are relevant for the relativistic interaction of localized wave packets, and we present a summary of our detailed investigation in the following. For a fixed $\chi=k_\perp \delta B / (k_\parallel B_0)$, a smaller ratio $k_\perp / k_\parallel$ corresponds to a larger $\delta B / B_0$ such that the system is driven away from the reduced MHD limit $\delta B_\perp / B_0 \sim k_\parallel / k_\perp \ll 1$ (Paper I). Hence, the amount of energy stored in the FW increases for a smaller ratio $k_\perp / k_\parallel$ (as we show in the top left panel of Fig.~\ref{fig:modes_packetsxi025}). During the initial phase of the interaction, fast waves in the $(1,1)$ mode show a progressive decay in energy after each primary wave collision. In contrast, $(1,1)$ Alfv\'{e}n waves incrementally increase in energy during each primary interaction event and conserve their energy in between. Fast waves effectively propagate as spherical waves in a periodic domain \citep[see also][]{SupplementaryMediaC,SupplementaryMediaH}. They, thus, continuously interact with Alfvén waves and other fast waves; energy is transferred out of the $(1,1)$ mode at a higher rate than for their Alfv\'{e}nic counterpart.} Fig.~\ref{fig:modes_packetsxi025} (right panel) probes the decomposition of the wave fields in the $(1,1)$ mode along the aforementioned characteristic directions. For the Alfv\'{e}n mode {(analyzed in the right middle panel of Fig.~\ref{fig:modes_packetsxi025})}, it is then straightforward to {identify} the field polarizations derived in Eq.~(\ref{eq:AWpolarization}). However, a quantitative analysis of the group velocity as well as the field polarizations of the fast wave (Eq.~\ref{eq:FWpolarization}) is more difficult. This intricacy is rooted in the finite spectrum of $k_\parallel$ due to the localization of the initial pulses, i.e., $k_\parallel$ is not a constant. A careful decomposition of wave modes for each individual $k_\parallel$ is required to confirm the remaining dispersion properties locally in Fourier space. We conducted such a brute-force validation. It shows that the $(1,1)$ mode is, indeed, the superposition of a linear Alfv\'{e}n wave and a linear fast wave. 

We note that in contrast to the continuous interaction of AWs, the localized $(1,1)$ mediator mode is a superposition of waves with a finite spectrum in $k_\parallel$. More specifically, it is \emph{not} a nonlinear fluctuation with $k_\parallel=0$. This distinction of spectral components is the key to understanding the localized ${\mathrm{AW}}+{\mathrm{AW}}$ interaction. Counter-propagating Alfv\'{e}n waves interact resonantly and satisfy the matching conditions
\begin{align}
    \mathbf{k}_1^++\mathbf{k}_1^-=\mathbf{k}_2,\qquad\omega_1^++\omega_1^-=\omega_2.
    \label{eq:AWresonance}
\end{align}
\begin{figure}
  \centering
  \includegraphics[width=1\textwidth]{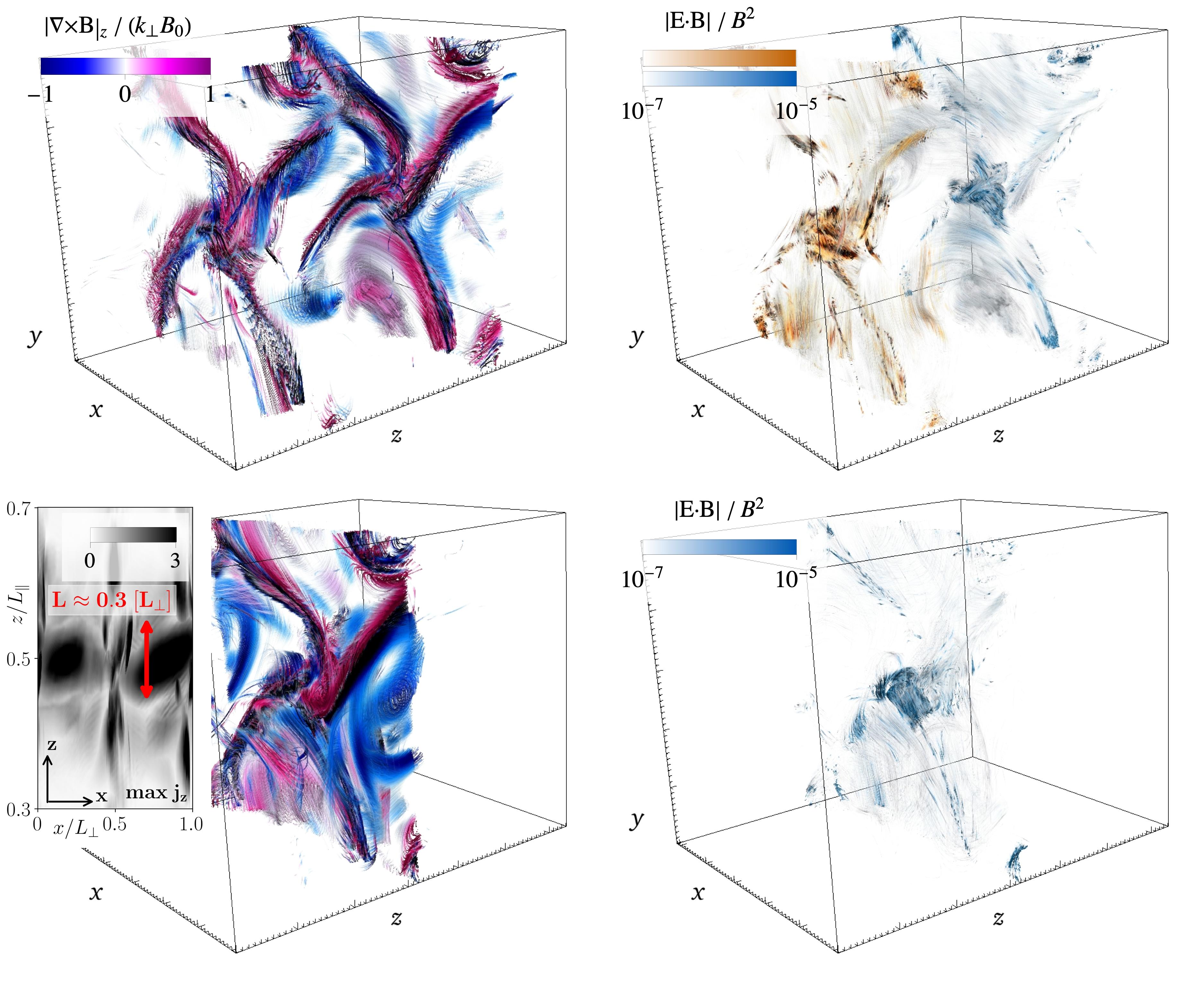}
    \vspace*{-22pt} 
    \begin{flushright}
    \textcolor{blue}{\texttt{ET-FFE}}
    \end{flushright}
  \caption{3D impressions of the ${\mathrm{AW}}_3+{\mathrm{AW}}_3$ collision during a phase of packet separation ({$t=31 [2\pi/\omega_0]$}, \textit{top row}) and the subsequent interaction ({$t=31.5 [2\pi/\omega_0]$,} \textit{bottom row}). We color field lines along $\mathbf{\hat{B}}$ by the current density $|\nabla \times \mathbf{B}|_z/(k_{\parallel}B_0)$ (\textit{left column}) and by the local non-ideal electric field marker $\mathbf{E}\cdot\mathbf{B}/\mathbf{B}^2$ (\textit{right column}). The vortex-like anchor points of the shear (cf. Fig.~\ref{fig:Packets2D}) are extended along the $z$-direction and coincide with regions of non-ideal diffusion. The inset in the \textit{bottom left} panel shows a top view of the maximum current density to estimate the depth of the current structures. {The different colors in the \textit{top right} panel distinguish between field lines as seeded in the two central planes of the displayed wave packets. A 3D animation of the dynamics in the interaction region can be found in \citet{SupplementaryMediaF}.}}
\label{fig:Packets3DC}
\end{figure}

\begin{figure}
  \centering
  \includegraphics[width=1\textwidth]{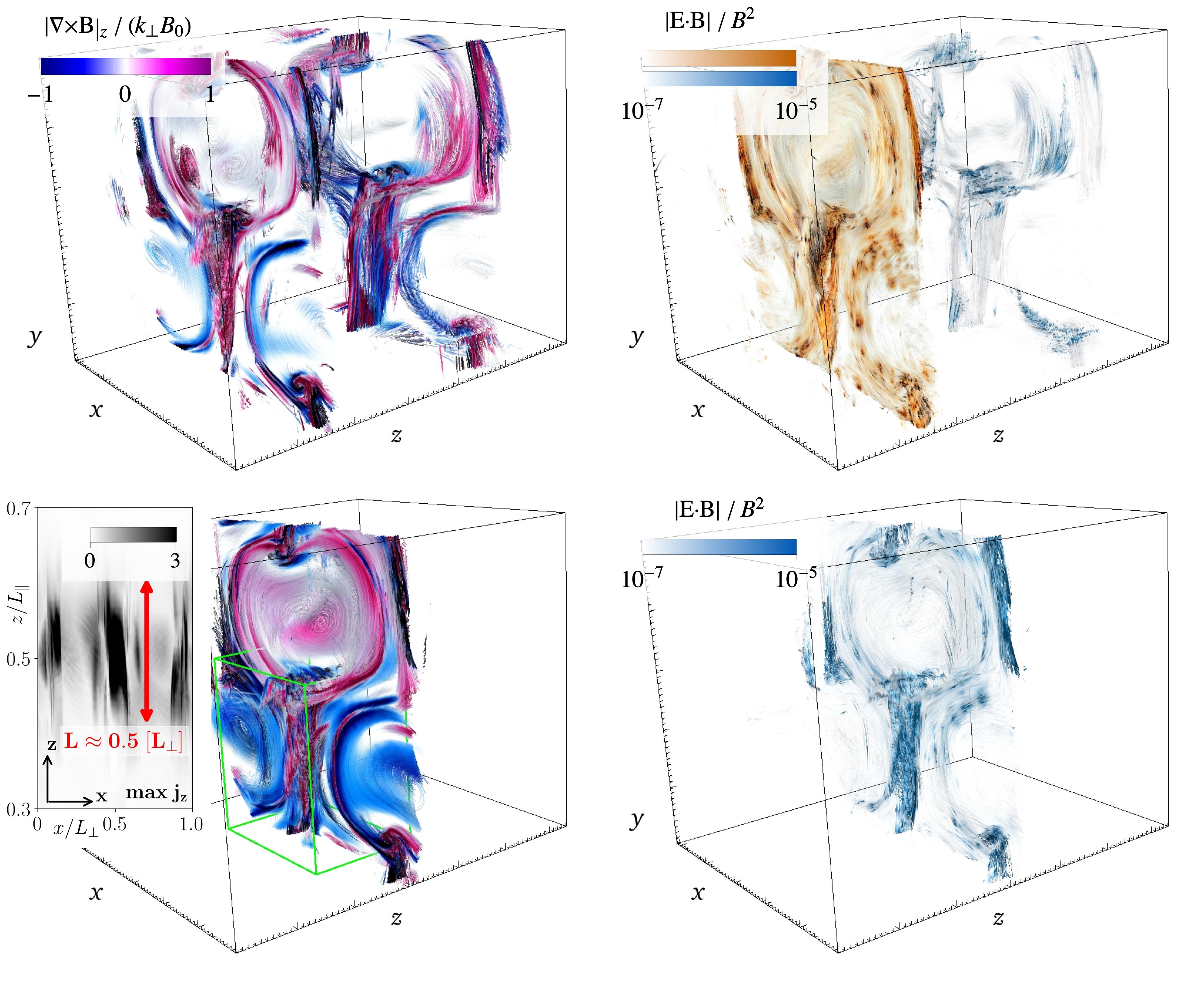}
  \vspace*{-22pt} 
    \begin{flushright}
    \textcolor{blue}{\texttt{ET-FFE}}
    \end{flushright}
  \caption{3D impressions of the ${\mathrm{AW}}_1+{\mathrm{AW}}_3$ collision in analogy to Fig.~\ref{fig:Packets3DC} {($t=31 [2\pi/\omega_0]$, \textit{top row}; $t=31.5 [2\pi/\omega_0]$, \textit{bottom row})}. A current layer separates two equally polarized eddies. It coincides with field reversals, strong currents, and significant dissipation by emerging non-ideal electric fields. A detailed outline of the reconnection region (green box) with realistic proportions is presented in Fig.~\ref{fig:Packets3DNULLS}. { A 3D animation of the dynamics in the interaction region can be found in \citet{SupplementaryMediaG}.}}
\label{fig:Packets3DOS}
\end{figure}

\begin{figure}
  \centering
  \includegraphics[width=1\textwidth]{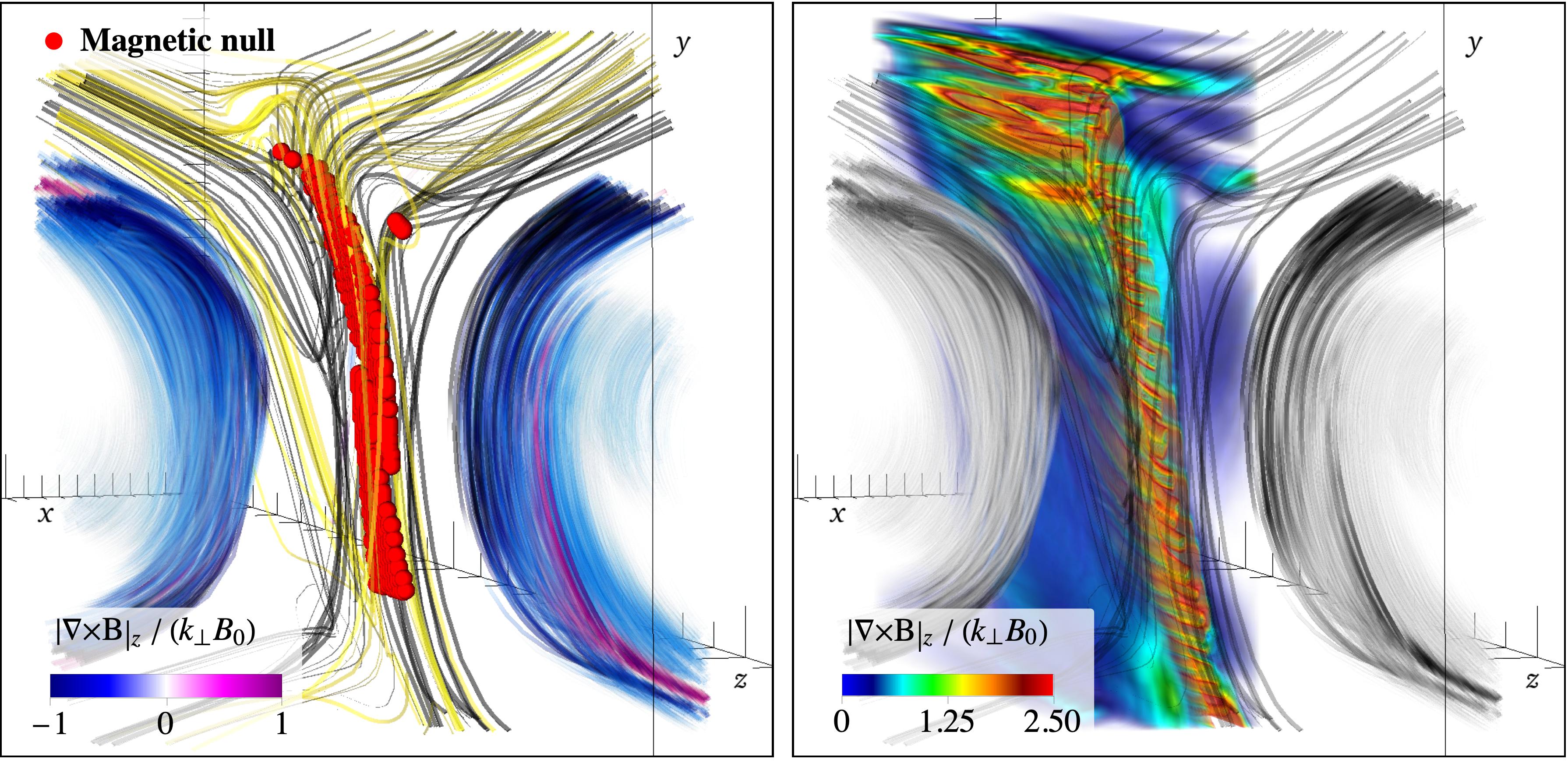}
  \vspace*{-11pt} 
    \begin{flushright}
    \textcolor{blue}{\texttt{ET-FFE}}
    \end{flushright}
  \caption{Detailed 3D outline of the reconnection region during the ${\mathrm{AW}}_1+{\mathrm{AW}}_3$ collision (as presented in Fig.~\ref{fig:Packets3DOS}). \textit{Left} panel: Magnetic null points (\textit{red} spheres, where we removed the initial background field $B_0$ along $\hat{\mathbf{z}}$) between eddies of the same polarity. \textit{Yellow} field lines are seeded at the locations of the strongest currents, \textit{black} field lines have a slight offset to these seeds in the perpendicular plane. \textit{Right} panel: Volume rendering of the strongest currents (as we showed for the continuous case in Fig.~\ref{fig:currentsheetFF512volume}).}
\label{fig:Packets3DNULLS}
\end{figure}

For these conditions to result in an outgoing AW as identified in Eq.~(\ref{eq:AWpolarization}), one of the incoming modes is required to have $k_\parallel=0$. Such a combination is possible when one wave-packet of a finite $k_\parallel$ spectrum interacts with the $k_\parallel=0$ component of another wave-packet \citep{Ng1996}. The resulting AW would then have a distribution of $k_\parallel$ that matches the primary mode; we find that this is indeed the case. An interaction between primary modes across the full spectral range of $k_\parallel$ also results in an outgoing FW. Due to the addition of spectral ranges, the FW should have an accumulation of power at $k_\parallel=0$. We observe this feature consistently in our analysis.

The interaction of \emph{continuous} counter-propagating AWs with the symmetry of constant and oppositely signed $k_\parallel$ has a unique solution to the resonance condition of Eq.~(\ref{eq:AWresonance}). Namely, it yields the nonlinear fluctuation (1,1,0). The \emph{localization} of waves in packets extends the spectrum of $k_\parallel$ along a finite range, and the symmetry in the resonance is broken. In contrast to the continuous interaction (Section \ref{sec:alfvendynamics}), the secondary mode has a linear Alfv\'{e}n wave character and propagates according to the dispersion relation given above.
During each interaction of the wave-packets, energy is transferred to the secondary (and higher order) modes. The secondary mode manifests via shearing of the magnetic field. While, initially, wave packets form elongated tubes along their respective $k_\perp$, the persistent shearing stretches these tubes, gradually compressing them (cf. Fig.~\ref{fig:Packets2D}). {The interaction of waves with symmetric $k_\parallel$ induces the (1,1,0) oscillating shear mode. Thus,} depending on the combination of initial $k_\parallel$, as to say, of the wave packets ${\mathrm{AW}}_1$ and ${\mathrm{AW}}_3$, the shearing motion can be continuously progressing (for non-symmetric $k_\parallel$) or oscillating (for symmetric $k_\parallel$).

In the case of ${\mathrm{AW}}_1+{\mathrm{AW}}_1$, the oscillation of the shearing process prevents the compression of anti-parallel field lines in between eddies into current sheets. However, elongated eddies, e.g., at $x\approx 3\pi/2, y\approx \pi/2 $ at $t=31\;[2\pi / \omega_0]$ in {Fig.~\ref{fig:Packets2D}\textbf{a}}, develop and form pancake-like current sheets, e.g., at $x\approx 4\pi/3, y\approx \pi/3 $ at $t=31\;[2\pi / \omega_0]$ in {Fig.~\ref{fig:Packets2D}\textbf{b}}. Prominent non-ideal electric fields form at these locations. The signatures of current sheets at later times ({Fig.~\ref{fig:Packets2D}\textbf{b}}) can be associated to a mild steepening of the energy dissipation. The shearing and stretching of larger eddies into reconnection layers are a characteristic of both continuous and localized Alfv\'{e}n wave interactions.

{In analogy to the case of continuous interaction (Section \ref{sec:alfvendynamics}), elongated eddies can form long current sheets that will eventually break up (cf. Section \ref{sec:currentsheetformation}).
As in the previous analysis of continuous interactions, we associate the onset of dissipation of cascaded energy with the emergence and break-up of current sheets. We identify sheets of strong current density, prominent non-ideal electric fields, and field reversals in the ${\mathrm{AW}}_1+{\mathrm{AW}}_3$ case ({Fig.~\ref{fig:Packets2D}\textbf{c}}) due to field line compression between two eddies at $x\approx \pi, y\in[0,\pi]$ in {Fig.~\ref{fig:Packets2D}\textbf{c}}. They emerge and decay at the same time as we observe a change in slope in the energy evolution.} 

We note that at the relatively high value of $\chi$ for wave packet simulations, while we have seen a persistence of the $E_{B_\perp} (k_{\perp})\propto k_{\perp}^{-2}$ spectrum despite the presence of intermittency (in the form of current sheets), we begin to see some features of a transition to strong turbulence (\citealt{Meyrand2016}) which call for further studies.

We conclude this section by studying 3D profiles of the wave interactions of the ${\mathrm{AW}}_3+{\mathrm{AW}}_3$ as well as the ${\mathrm{AW}}_1+{\mathrm{AW}}_3$ collision in Figs.~\ref{fig:Packets3DC} and~\ref{fig:Packets3DOS}. In these figures, we color a large number of field lines seeded in planes at the center of the respective wave packets (or their superposition) by the current density $|\nabla \times \mathbf{B}|_z/(k_{\parallel}B_0)$ (\textit{left column}) and by the local non-ideal electric field marker $\mathbf{E}\cdot\mathbf{B}/\mathbf{B}^2$ (\textit{right column}). First, we note that the wave magnetic field is well localized even after several tens of interactions. Second, possible 3D reconnection sites emerge in the form of bulged eddies (Fig.~\ref{fig:Packets3DC}) or current sheets with varying depth (Fig.~\ref{fig:Packets3DOS}). These structures exist individually in each wave packet and interact progressively, enhanced by the shearing action of the secondary mode. The $z$-extensions of non-ideal structures is a fraction of the initial window-size $\Delta_z$ in the case of the bulged eddies. Current sheet structures, however, have a significant depth of the order of (and even slightly above) the initial window-size $\Delta_z$. Fig.~\ref{fig:Packets3DNULLS} dissects the topology of the current sheet emerging during the ${\mathrm{AW}}_1+{\mathrm{AW}}_3$ collision, showing some anti-parallel field lines and some reconnecting field lines.
We mark the localized layer of magnetic null points with red dots (i.e. points with reversals of all field components, where the initial background field $B_0$ along $\hat{\mathbf{z}}$ is removed).

{Collisions of localized Alfv\'{e}n waves are efficient mediators of energy conversion to smaller scales (perpendicular to the guide field) and powerful drivers of nonlinear energy cascades.} The mixing of wave packets with different ratios of $k_\perp/k_\parallel$ significantly changes the interaction dynamics {by suppressing the oscillatory (1,1,0) shear mode. Mixing effectively} triggers episodes of very fast dissipation of electromagnetic energy mediated by the break-up of elongated current sheets. As in the continuous case (cf. Fig.~\ref{fig:currentsheetFF512volume}), the current sheet ripples as it is stretched between elongated eddies and eventually breaks up into smaller (turbulent) structures. {Mixing waves of different $k_\perp/k_\parallel$ is an appealing case for a future exploration of current sheet formation and the transition into the strong turbulence regime for the interaction of overlapping Alfv\'{e}n waves in high-resolution runs.}

\section{Conclusions}
\label{sec:conclusions}

In this work, we show that the interaction of {overlapping and counter-propagating Alfv\'{e}n waves in highly magnetized relativistic plasma} results in an anisotropic energy spectrum $E_{B_{\perp}}(k_{\perp}) \propto k_{\perp}^{-2}$. 
The weak turbulence energy cascade occurs due to the scattering of Alfv\'{e}n waves dominated by three-wave interactions.
The nonlinear dynamics of Alfv\'{e}n wave collisions provide a natural mechanism for the development of current sheets in plasma turbulence.
Turbulent eddies that form through constructive interference of the primary Alfv\'{e}n waves and nonlinearly generated modes become elongated and stretched on nonlinear time scales $\sim \chi^{-2}$, until they transiently form current sheets. The current sheets undergo a thinning process due to the compression of the eddies, resulting in a final thickness $\delta / {w} \approx 0.01$ before they break-up into small-scale turbulent structures. 
Analyzing the electromagnetic energy evolution of the system, we find that the turbulent cascade reaches the grid scale at the moment the current sheets break-up, at which time {the} energy dissipates dramatically. We {suggest} that magnetic reconnection in the current sheet regions is a viable mechanism through which energy dissipates in these highly relativistic and magnetically dominated fluid systems.

{The presented results show} that the fundamental properties of Alfv\'{e}n wave collisions as observed in the idealized case of periodic and overlapping waves  persist under the more realistic conditions of localized wave packets. {The} evolution and interaction of the wave packets occurs during their overlap only, and the {Alfv\'{e}n component of the }wave packets remains localized along the guide magnetic field before and after their collision. {As a consequence,} packet collisions are in many ways similar to the continuous wave collisions. 
{Especially}, strong dissipation {of electromagnetic energy} in current sheets starts to develop at nonlinear timescales of order $\chi^{-2}$. {In contrast to the continuously overlapping waves, the} mediator of the turbulent cascade that transfers energy to the smallest scales is not a purely nonlinear magnetic mode, but a combination of linear Alfv\'{e}n and fast waves. \

{In fact, fast waves are absent in the non-relativistic limit \citep{Verniero_2018,Verniero_2018b}; thus, despite a multitude of similarities between Alfv{\' e}nic turbulence in the Newtonian versus relativistic limits, this notable difference provides at least one alternative path for the turbulence to dissipate its energy since the linear fast modes can travel across field lines. Regardless, as in the Newtonian limit, the secondary Alfv{\' e}n mode is essentially a shear in the magnetic field that propagates along the guide field and shears the counter-propagating Alfv{\' e}n wave packets. In addition, for a fixed $\chi=k_\perp \delta B / (k_\parallel B_0)$, a large ratio of $k_\perp/k_\parallel$ results in a smaller amount of energy stored in the fast waves, in accordance with the asymptotic solutions in Paper I.} We have demonstrated that current sheets form during the interaction of two packets {for different ratios of $k_\perp/k_\parallel$} and that they act as the main dissipation sites.

Although the force-free limit of the MHD equations is technically invalid inside a reconnecting current sheet (where $E>B$ or $\mathbf{E} \cdot \mathbf{B} \neq 0$) and it cannot describe the physical effects of magnetic reconnection and resistive dissipation, it does contain the minimal ingredients that lead to the anisotropic cascade and the development and thinning of current sheets in relativistic plasma turbulence. We have validated our results for finite magnetization in magnetically dominated ideal relativistic MHD. We {suggest} that the emergence and dynamic decay of {current sheets} is driven by the global dynamics of wave interactions. {We verified that at the highest presented resolutions, }the small structures we identify are not dominated by numerical diffusion or dispersion errors {(cf. App.~\ref{sec:appendixa})}. To validate our findings, we emphasize{, once more,} the striking {similarity} of results from two independent{, vastly different} force-free MHD algorithms. To study current sheets as dissipation sites in interacting Alfv\'{e}n waves in more depth, we will in the future carry out relativistic resistive MHD simulations \citep{Ripperda_2019,Ripperda_2019a} and evolve test particles to capture magnetic reconnection as a particle heating and acceleration mechanism \citep{Ripperda_2017a,Ripperda_2017b,Ripperda_2018}.

{Though obtained in the context of fundamental plasma physics, the results are vital for the magnetospheres of astrophysical compact objects.} Turbulence in black hole accretion disks may launch Alfv\'{e}n waves that propagate from the disk into the corona \citep{Thompson_1998,Chandran_2018}. {Such} waves may propagate away from the disk into regions with varying magnetization and reflect, after which the interaction with other waves may result in a turbulent cascade. {At this point,} a significant fraction of the wave energy can dissipate within a few scale heights of the disk. The current sheets formed by relativistic Alfv\'{e}nic turbulence as found in this work may provide the main dissipation sites of magnetic energy through magnetic reconnection, yielding a promising mechanism for explaining the X-ray emitting coronae that are observed around luminous active galactic nuclei. Even in the weak turbulence regime ($\chi < 1$) explored here, a large reservoir of magnetic energy is available in the turbulent fluctuations $\delta B_{\perp}$ in magnetized accretion disk coronae. The turbulence in black hole accretion disks may even be in the strong turbulence limit (i.e., $\delta B_{\perp}/B \geq 1$), a regime we {explore in another study (\citealt{Chernoglazov2021})}.

{The interaction of wave packets is equally relevant in neutron star magnetospheres.} Magnetar flares can excite strong Alfv\'{e}n waves in the highly magnetized magnetosphere. The wave packets can dissipate their energy through interactions with reflecting waves \citep{Li_2019}, {or they} may dissipate their energy in the neutron star's crust \citep{Li_2015}. They can {also} convert to fast modes which are not confined to the magnetic field and can escape from the magnetosphere \citep[{cf.}][]{yuan2020alfvn}. {A} turbulent cascade of large amplitude Alfv\'{e}n waves in the magnetosphere can lead to plasma heating and X-ray emission. The {fraction and rate of }dissipation constrains the duration and flux {of the observed X-ray emission} following {a} magnetar flare. {On another note}, pulsar glitches may launch Alfv\'{e}n waves into the magnetosphere, which {can} lead to enhanced current and pair production that quenches the radio emission \citep{Bransgrove_2020,Yuan_2020}. The lifetime, dissipation channel, and the ratio between turbulent Alfv\'{e}n and fast modes (dependent on the $k_\parallel / k_\perp$ ratio of the primary waves{, as we show in Section \ref{sec:alfvendynamics}}) constrains the duration of the observed radio emission anomaly. We {argue} in this work that it is essential to study wave interaction in neutron star magnetospheres in {three dimensions} to determine the fraction of energy that can escape through fast modes {in future magnetospheric modeling.}

{Finally, looking beyond the validity of our models, we note that the} low density of the plasma in the magnetospheres of neutron stars and black holes results in a mean free path for collisions that is much larger than the length scales of turbulent fluctuations, often even exceeding the system size. {Thus, plasma can be considered collisionless, and turbulent dynamics and dissipation is governed by kinetic physics.} The MHD approximation fails in this regime and cannot capture the collisionless kinetic physics, {potentially resulting} in non-thermal radiation that is typically observed. {Throughout the results presented in this manuscript, we observe strong guide field current sheets emerging as a result of Alfv\'{e}n wave collisions. In this regime,} the energy available for plasma energization is limited, resulting in {significantly steeper} power-law spectra {shown by the first-principles particle-in-cell simulations of} \cite{Werner_2017}. {By using gyrokinetic simulations, \citet{TenBarge_2013} and \citet{Howes_2018}} showed that plasma heating is indeed correlated with the emergence of current sheets. Landau damping plays an essential role in the spatially intermittent energization of particles through dissipation of turbulent fluctuations. Understanding the mechanism that energizes the plasma in black hole accretion disk coronae and neutron star magnetospheres is an important problem in high-energy astrophysics. {Force-free and relativisic MHD simulations capture the essential global dynamics to describe a weak turbulence cascade and the formation of current sheets that can act as dissipative regions in highly magnetized plasma as found in compact object magnetospheres and coronae.}


\section*{Acknowledgements}

We acknowledge the Flatiron's Center for Computational Astrophysics and the Princeton Plasma Physics Laboratory for support of collaborative CCA-PPPL meetings on plasma-astrophysics where the ideas presented in this paper have been initiated.
The computational resources and services used in this work were provided by facilities supported by the Scientific Computing Core at the Flatiron Institute, a division of the Simons Foundation; and by the VSC (Flemish Supercomputer Center), funded by the Research Foundation Flanders (FWO) and the Flemish Government – department EWI. {The calibration of \textsc{ET-FFE} was in part conducted on resources of the Barcelona Supercomputing Center (AECT-2021-1-0006).}
BR is supported by a Joint Princeton/Flatiron Postdoctoral Fellowship. AAP and JFM acknowledge support by the National Science Foundation under Grant No. AST-1909458. ERM gratefully acknowledges support from a joint fellowship at the Princeton Center for Theoretical Science, the Princeton Gravity Initiative and the Institute for Advanced Study. JMT and AB acknowledge support by the Simons Foundation. AC gratefully acknowledges support and hospitality from the Simons Foundation through the pre-doctoral program at the Center for
Computational Astrophysics, Flatiron Institute. JJ is supported by a NSF Atmospheric and Geospace Science Postdoctoral Fellowship (Grant No. AGS-2019828). YY is supported
by a Flatiron Research Fellowship at the Flatiron Institute, Simons Foundation. Research at the Flatiron Institute is supported by the Simons Foundation. We thank Gregory Howes, Joonas N\"{a}ttil\"{a}, Andrei Beloborodov, Daniel Groselj, Fabio Bacchini, and Lev Arzamasskiy for useful discussions.
BR and JFM contributed equally to this work.
Competing interests: The author(s) declare none.


\appendix
\section{Numerical convergence of the force-free MHD schemes}
\label{sec:appendixa}

We test our newly implemented force-free algorithm in {\tt BHAC} for an Alfv\'{e}n wave propagating in the $z$-direction along the guide field $B_z = B_0$ with wave vector $\mathbf{k}=(1,0,1)$, for 200 wave-crossings through a box of size $(2\pi,2\pi,4\pi)$ with $N_{\lambda,z}$ cells per wavelength and resolution $(N_{\perp},N_{\perp},N_{\parallel})$ where $N_{\lambda,z} = N_{\perp} = 0.5N_{\parallel} \in [8,16,32,64,128,256,512]$. We measure the numerical energy dissipation by plotting the maximum (taken at every point in time over the whole domain, i.e., not necessarily at the same location) of the electric energy density $\max[E^2]$, which should be conserved in ideal force-free MHD for a non-interacting wave. 

\begin{figure}
  \centering
  \includegraphics[width=1.0\textwidth]{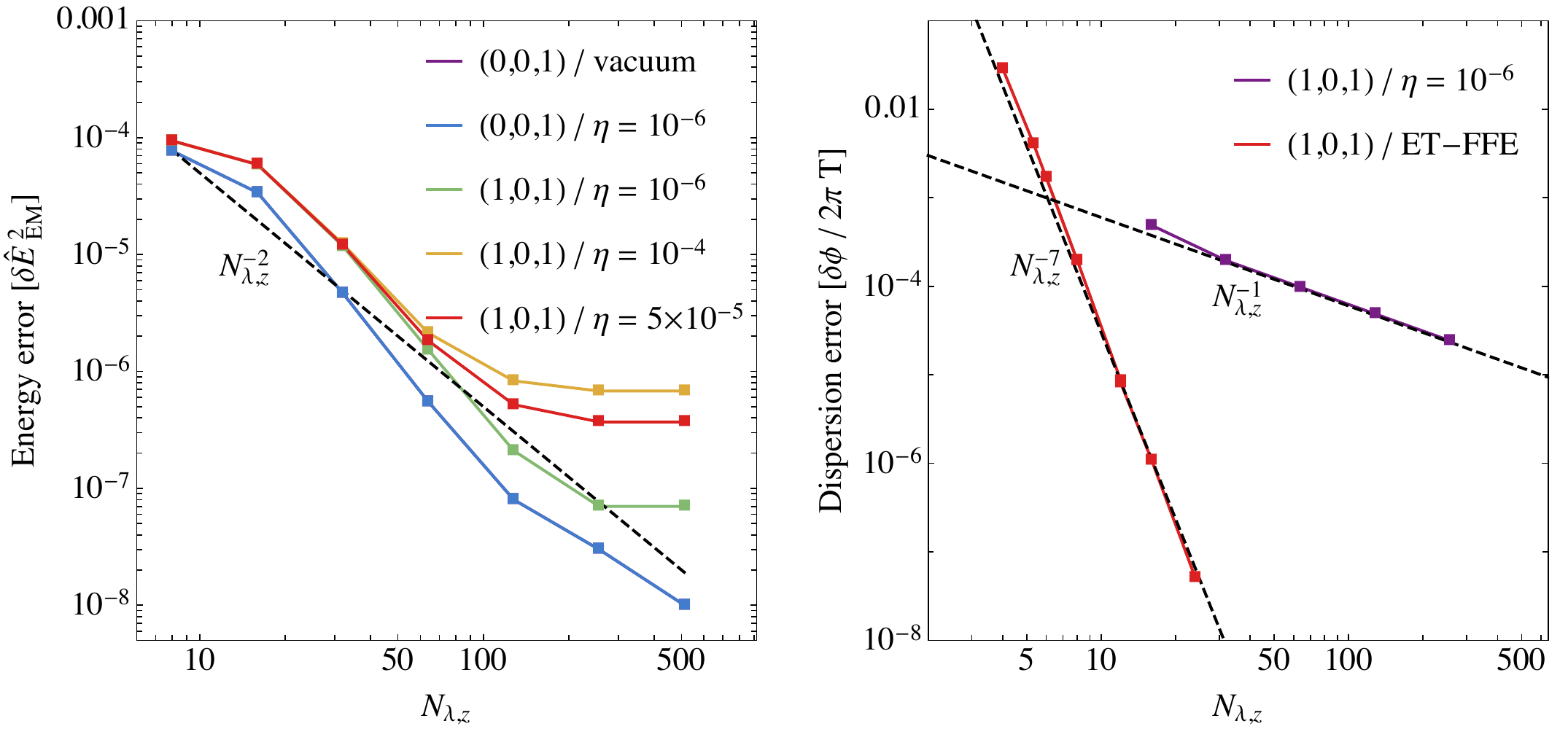}
  \vspace*{-22pt} 
    \begin{flushright}
    \textcolor{blue}{\texttt{BHAC}/\texttt{ET-FFE}}
    \end{flushright}
  \caption{{Convergence properties of the employed numerical methods. }Left: Second order convergence of the electric energy density error $\delta {\hat{E}_{\rm EM}^2} = ||\max[{\hat{E}_{\rm EM}^2}]_{\omega_0 t / 2\pi=10} - \max[{\hat{E}_{\rm EM}^2}]_{\omega_0 t / 2\pi=0}||$ versus resolution per wavelength $N_{\lambda,z}$, for single waves with $\mathbf{k}=(0,0,1)$ in force-free MHD with $\eta=10^{-6}$ (the force-free limit), and the electrovacuum limit, and for a single wave with $\mathbf{k}=(1,0,1)$ with $\eta=10^{-6}$. For the $\mathbf{k}=(1,0,1)$ wave, we find that the explicit resistivity dominates the dissipation from $N_{\lambda,z}=256$ onward, and the error saturates. For the $\mathbf{k}=(0,0,1)$ wave, we {find} that the error in the energy conservation is independent of the effective resistivity $\eta$ as expected for a wave that has zero conduction current. We find second-order convergence, as expected for our IMEX scheme, both in the ideal force-free MHD limit $\eta \rightarrow 0$ and in the electrovacuum limit $\eta \rightarrow \infty$ in Eq. \ref{eq:ohmslaw}. For $N_{\lambda,z}=8$ cells per wavelength, the error is equal to the initial wave energy since most energy is dissipated after 10 collisions. We confirm that the error is dominated by the spatial resolution and that it does not decrease for a decreasing time step at a given resolution of $N_{\lambda,z}=16$. Right: The dispersion error of a single wave as the phase shift $\delta \phi = \delta x / 2\pi$ of the nulls of the wave normalized by the number of wave periods $T = t / 2\pi$. The error converges as first order $N_{\lambda,z}^{-1}$ with grid points per wavelength and is dominated by the temporal error that is first order due to the implicit step in the IMEX scheme for {\tt BHAC}, whereas it converges at seventh order for the {\tt ET-FFE} code.}
\label{fig:singlewaveenergy}
\end{figure}

We analyze the error in the electric energy density because the dissipation operator appears in Ampere's law (Eq. \ref{eq:ampereFF}). The electromagnetic energy at $N_{\lambda,z} = 8$ cells per wavelength has completely dissipated after 10 collisions and the error is equal to the initial energy in the wave. By increasing the resolution per wavelength, we find second-order convergence of the error $\delta E^2 = ||\max[E^2]_{\omega_0 t / 2\pi=10} - \max[E^2]_{\omega_0 t / 2\pi=0}||$ after ten wave-crossings, both in the ideal force-free MHD limit ($\eta \rightarrow 0$ in Eq. \ref{eq:ohmslaw}), by setting $\eta=10^{-7}$, and in the electrovacuum limit ($\eta\rightarrow\infty$ in Eq. \ref{eq:ohmslaw}), by setting $\eta=10^{14}$ (see Fig. \ref{fig:singlewaveenergy}). We confirm that the error in the energy conservation is independent of the effective resistivity for $\eta \in [10^{-3},10^{-4},10^{-5},10^{-6},10^{-7}]$. The specific resistivity sets a resistive spatial scale that cuts off the energy error convergence once it reaches that scale, thus, the error becomes smaller for smaller resistivity. We compare the result with a wave vector $\mathbf{k}=(0,0,1)$ that is aligned with the guide field, such that there is no damping term in Eq. \ref{eq:ohmslaw} and, hence, the resistivity has no effect on the energy dissipation. There, we observe that the energy error converges at second order and there is indeed no cut-off at the resistive scale. The convergence is dominated by the spatial order of the scheme, which we {validate} by measuring the error for a decreasing time step at a fixed resolution of $N_{\lambda,z}=16$. A similar analysis was conducted for the \texttt{ET-FFE} algorithm \citep{mahlmann2020computationala}, showing convergence close to the spatial order of reconstruction, i.e., seventh-order in the simulations presented throughout this manuscript.

Furthermore, we determine the dispersion error as the shift $\delta \phi = \delta x / 2\pi$ of the nulls of the wave normalized by the number of wave periods $T = t / 2\pi$. We measure the error after 1, 10 and 20 wave-crossings and summarize the results in the right panel of Fig.~\ref{fig:singlewaveenergy}. The dispersion error is dominated by temporal errors and converges at first order due to the implicit step that makes the IMEX scheme first order in time (\citealt{Ripperda_2019}). We confirm that the dispersion error linearly increases in time such that we can normalize the results by the wave-period. With a seventh order spatial reconstruction and fourth order time integration, the dispersion error of the \texttt{ET-FFE} algorithm decreases slightly faster than the order of reconstruction. In other words, this method is more diffusive then it is dispersive and, hence, well suited for the long time modeling of wave interactions.

\section{Non-ideal electric fields in (force-free) current sheets}
\label{sec:appendixb}

\begin{figure}
  \centering
  \includegraphics[width=1.0\textwidth]{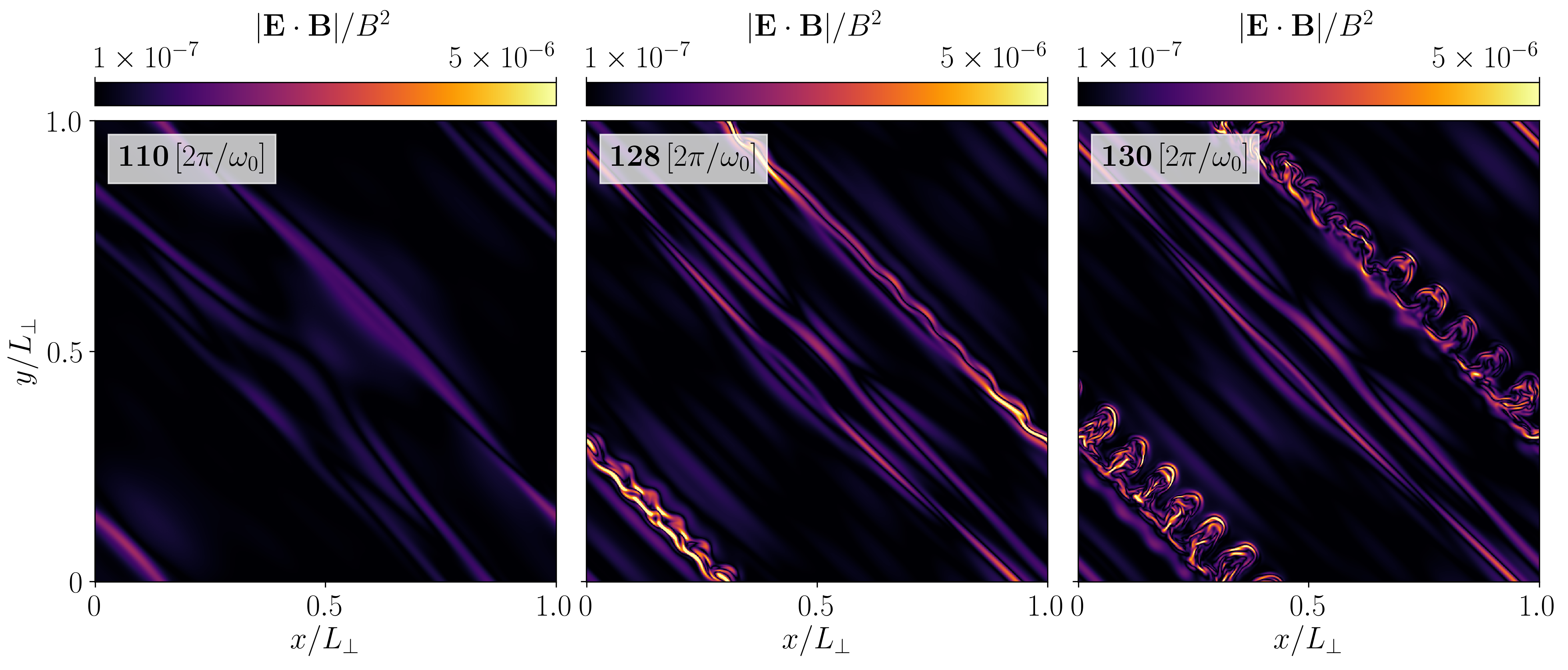}
  \vspace*{-22pt} 
    \begin{flushright}
    \textcolor{blue}{\texttt{BHAC}}
    \end{flushright}
  \caption{Non-ideal electric field $|\mathbf{E} \cdot \mathbf{B}| / B^2$ growing inside the current sheets. We examine such fields shortly after the formation of the sheets at $t=110\;[2\pi/\omega_0]$, at the end of the thinning period of the sheets to $\delta/w\approx0.01$ at $t=128\;[2\pi/\omega_0]$, and at their break-up around $t=130\;[2\pi/\omega_0]$. Here, we employ the force-free MHD infrastructure in {\tt BHAC} with $\eta=10^{-6}$. The violations of the ideal force-free condition are damped by resistive dissipation (see Section \ref{sec:numerics}).}
\label{fig:BHAC_EdotB}
\end{figure}

\begin{figure}
  \centering
  \includegraphics[width=1.0\textwidth]{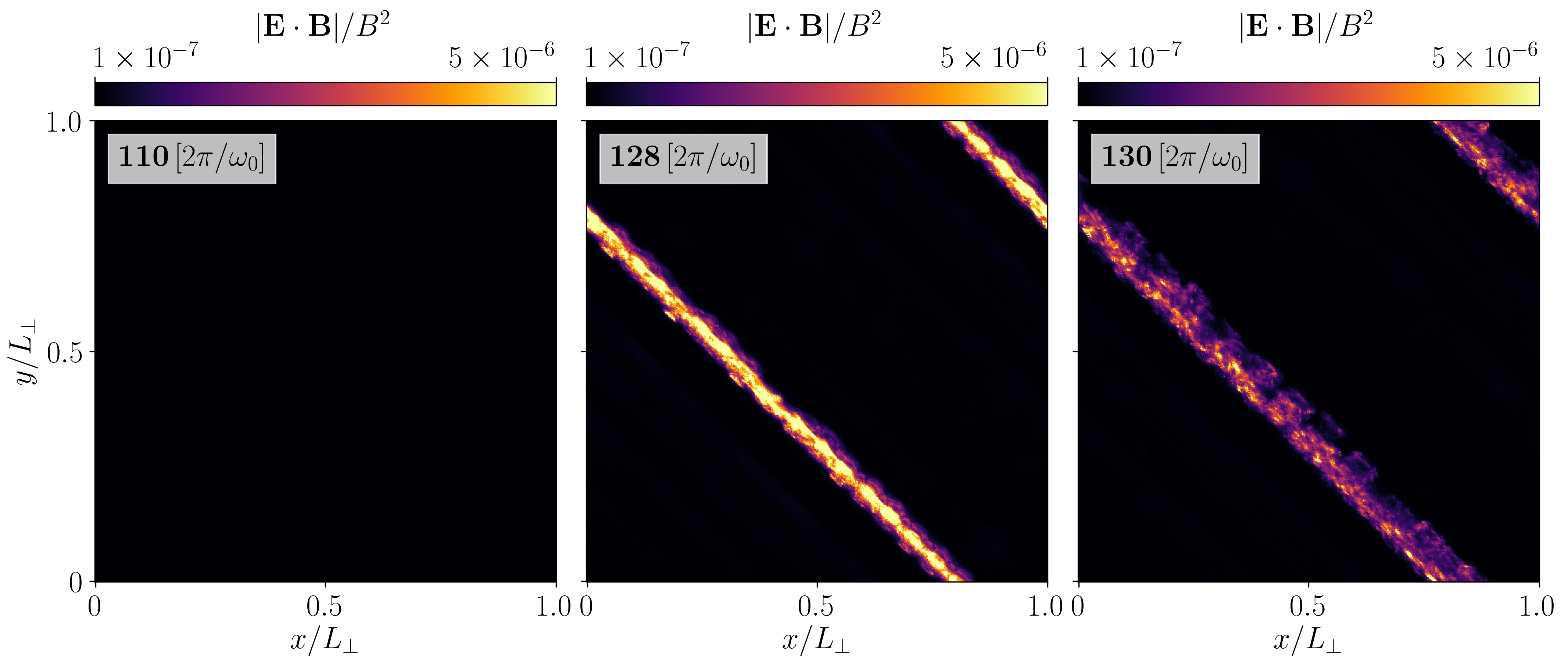}
  \vspace*{-22pt} 
    \begin{flushright}
    \textcolor{blue}{\texttt{ET-FFE}}
    \end{flushright}
  \caption{As Fig.~\ref{fig:BHAC_EdotB} but displaying the magnitude of the non-ideal electric field (averaged along the $z$-directions) that is removed in each sub-step of the time-integrator at a given time by the \texttt{ET-FFE} scheme. We note the similarities to the non-ideal electric field that is allowed for finite times by \texttt{BHAC} (Fig.~\ref{fig:BHAC_EdotB}).}
\label{fig:ET_EdotB}
\end{figure}

Current sheets, i.e., regions of anti-parallel magnetic field where typically $\mathbf{E}\cdot\mathbf{B}\neq0$ or $E^2>B^2$, are inherently difficult to model in the force-free limit of infinite conductivity. Non-ideal electric fields, characterized by $\mathbf{E}\cdot\mathbf{B}\neq0$, can form as a result of magnetic reconnection in a current sheet. Force-free evolution codes either allow for transient non-ideal electric fields and gradually damp them (as is the case in \texttt{BHAC}, cf. Eq.~\ref{eq:ohmslaw}), or they remove them instantly from the domain in each sub-step of the time-integrator (as it is the case in \texttt{ET-FFE}).

Figs. \ref{fig:BHAC_EdotB} and~\ref{fig:ET_EdotB} show the non-ideal electric fields accumulating in the different frameworks. In the case of the \texttt{ET-FFE} data, a direct measurement of $E_\parallel$ (or equally, of an Ohmic heating term) is not possible, as all violations of the force-free conditions are instantly removed from the domain. Instead, we track the magnitude of parallel electric fields that are algebraically removed in each sub-step of the time integrator, namely $\mathbf{E}\cdot\mathbf{B}/B^2$. In both cases, we find direct coincidence of the location of field reversals and other markers for current sheets with regions of stronger $E_\parallel$. Furthermore, such fields are still small at $t=110\;[2\pi/\omega_0]$, but they grow significantly during the thinning and break-up phases of the sheets.

\texttt{BHAC} and \texttt{ET-FFE} differ significantly in their treatment of non-ideal electric fields. This distinction is imprinted on the data by the emergence of small-scale structures that are visible in Figs.~\ref{fig:BHAC_EdotB} and~\ref{fig:ET_EdotB}. The combination of a finite phenomenological resistivity and lower-order spatial reconstruction in \texttt{BHAC} renders strong $E_\parallel$ significant for the resolution of thin resistive layers occurring in current sheets. \texttt{ET-FFE} deals with force-free violations much more rigorously, removing even the smallest $E_\parallel$ instantly and preventing the physical development of the smallest resistive layers \citep[but increasing the overall order of convergence; see extensive discussion in][]{mahlmann2020computational}. The fact that despite these differences, our results across the platforms \texttt{BHAC} and \texttt{ET-FFE} show a remarkable level of similarity reassures us of the physical validity of the conclusions presented in this manuscript.

\bibliographystyle{jpp}

\bibliography{jpp-instructions}

\end{document}